\documentclass[useAMS,usenatbib]{mn2e}

\usepackage{amsmath}             
\usepackage{graphicx}
\usepackage{txfonts}

\usepackage{natbib}

\newcommand{\msun}{~{\rm M}_\odot}
\newcommand{\kms}{km\,s$^{-1}$}

\newcommand{\amed}{a_{\rm med}}
\newcommand{\halfmass}{R_{\rm hm}}

\newcommand{\mcl}{M_{\rm cl}}
\newcommand{\ncl}{N_{\rm cl}}

\newcommand{\binfrac}{{\mathcal B}}
\newcommand{\nssf}{{\mathcal N}}
\newcommand{\csf}{{\mathcal C}}

\newcommand{\dynamicalpeak}{{\em dynamical peak}}
\newcommand{\dissolutionpeak}{{\em dissolution peak}}

\title[The formation of very wide binaries during the star cluster dissolution phase]{The formation of very wide binaries during the star cluster dissolution phase}
\author[M. B. N. Kouwenhoven, S. P. Goodwin, R. J. Parker,
M. B. Davies, D. Malmberg and
P. Kroupa]{M. B. N. Kouwenhoven$^{1,2}$\thanks{E-mail: kouwenhoven@kiaa.pku.edu.cn;
    s.goodwin@sheffield.ac.uk; r.parker@sheffield.ac.uk;
    mbd@astro.lu.se, danielm@astro.lu.se,
    pavel@astro.uni-bonn.de}\thanks{Peter and Patricia Gruber
    Foundation Fellow},
  S. P. Goodwin$^{2}$, Richard J. Parker$^{2}$, M. B. Davies$^{3}$,
  \newauthor D. Malmberg$^{3}$ and P. Kroupa$^{4}$\\
$^{1}$Kavli Institute for Astronomy and Astrophysics at Peking
University, Yi He Yuan Lu 5, Hai Dian District, Beijing 100871, P.R. China\\
$^{2}$University of Sheffield, Hicks Building, Hounsfield Road, Sheffield S3~7RH, United Kingdom\\
$^{3}$Lund Observatory, Box 43, SE-221~00, Lund, Sweden  \\
$^{4}$Argelander Institute for Astronomy, University of Bonn, Auf dem H\"{u}gel 71, 53121 Bonn, Germany}
\begin{document}

\date{Accepted ---. Received ---; in original form ---}

\pagerange{\pageref{firstpage}--\pageref{lastpage}} \pubyear{2008}

\maketitle

\label{firstpage}


\begin{abstract}
  Over the past few decades, numerous wide ($>10^3$~au) binaries in the
  Galactic field and halo have been discovered. Their existence cannot
  be explained by the process of star formation or by dynamical
  interactions in the field, and their origin has long been a
  mystery. We explain the origin of these wide binaries by formation
  during the dissolution phase of young star clusters: an initially
  unbound pair of stars may form a binary when their distance in
  phase-space is small. Using $N$-body simulations, we find that the
  resulting wide binary fraction in the semi-major axis range
  $10^3~\mbox{au} < a < 0.1~\mbox{pc}$ for individual clusters is
  $1-30\%$, depending on the initial conditions. The existence of
  numerous wide binaries in the field is consistent with observational
  evidence that most clusters start out with a large degree of
  substructure. The wide binary fraction decreases strongly with
  increasing cluster mass, and the semi-major axis of the newly formed
  binaries is determined by the initial cluster size. The resulting
  eccentricity distribution is thermal, and the mass ratio
  distribution is consistent with gravitationally-focused random
  pairing. As a large fraction of the stars form in primordial
  binaries, we predict that a large number of the observed ``wide
  binaries'' are in fact triple or quadruple systems. By integrating
  over the initial cluster mass distribution, we predict a binary
  fraction of a few per cent in the semi-major axis range
  $10^3~\mbox{au} < a < 0.1~\mbox{pc}$ in the Galactic field, which is
  smaller than the observed wide binary fraction. However, this
  discrepancy may be solved when we consider a broad range of cluster
  morphologies.
\end{abstract}



\section{Introduction} \label{section:intro}

A significant fraction of stars in the Galactic field are in binary
and multiple systems \citep[e.g,][]{duquennoy1991, fischer1992,
  mason1998,shatsky2002, goodwinkroupa2005, kobulnicky2007,
  kouwenhoven_adonis, kouwenhoven_recovery, lada2006,
  zinneckerreview2007,goodwinpp2007}.  It is also thought that the
majority of stars are born in star clusters
\citep{ladalada}. Therefore the majority of binaries\footnote{For
  brevity we will use `binaries' to mean `multiples' of any
  multiplicity for the remainder of this paper, only drawing a
  distinction where it is necessary.} in the field population
presumably originate from clustered star formation.

It is well known that binaries are dynamically processed in star
clusters with wider and less bound systems tending to be destroyed by
encounters \citep{heggie1975, hills1975}.  Therefore, the field binary
population is dynamically processed with respect to the birth
population of binaries \citep{kroupa1995b, parker2009}. The origin of
most field binaries can be understood as a mixture of differently
processed initial populations \citep{goodwin2009}.

However, a significant number of very wide ($a > 10^3$~au) binaries
have been observed in the field (see \S~\ref{section:howmany}).  As such wide
binaries are extremely sensitive to destruction they have been used to
constrain the properties of the Milky Way. Wide binaries in the Galactic
disc and halo have been used to place limits on the density of MACHOs
and other unseen material \citep[e.g,.][]{bahcall1985,quinn2009}, to
constrain the formation history of the Galaxy
\citep[e.g.,][]{allenpoveda2007}, and to test the dark matter
hypothesis \citep[e.g.,][]{hernandez2008}.  But the extreme
sensitivity to destruction makes the survival (and even formation) of
such binaries in a cluster something of a mystery.

Many very wide binaries have separations comparable to the average
interstellar separation in clusters (typically a few $10^3$~au), and
the very widest binaries have separations of order the size of a young
cluster core (typically a few $10^4$~au). Given this, it is difficult
to see how they could even form, let alone survive, in a cluster
\citep[see, e,g.,][]{scally1999, parker2009}. Even in an isolated
star forming region the typical size of a star forming core is only
$10^4$~au \citep{ward2007} which presumably sets the very maximum size
of a (primordial) binary system.

It is possible that a wide binary forms via dynamical interactions in
the Galactic field (the capture mechanism). A prerequisite for this
mechanism to work is that a significant amount of kinetic energy is
dissipated. This energy dissipation can occur due to tidal friction
and due to three-body interaction. Tidal friction occurs in the rare
event of two stars nearly colliding in a close encounter. In the vast
majority of the cases this results in a merger or a fly-by, and only
in a small number of cases does this lead to the formation of a binary
system. However, all binaries resulting from capture by tidal friction
are very tight, with orbital period of several days.

Another possible mechanism is three-body interactions. In this 
case the third star acts as the energy sink, and is
generally ejected with high velocity. However, the stellar density in
the field is low, of order $0.1 \msun\,{\rm pc}^{-3}$, so that
three-body encounters are rare, and capture by dynamical friction
rarely occurs. \cite{goodman1993} find that the creation rate
$\dot{N}_B$ for binaries per unit volume can be approximated by
\begin{equation} \label{equation:hut}
  \dot{N}_B = 0.75 \, \frac{G^5 M^5 n^3}{\sigma^9_v} \,,
\end{equation}
where $M$ is the typical mass of a star in the field, $n$ the number
density of stars, $\sigma_v$ the velocity dispersion, and $G$ the
gravitational constant. In the solar neighbourhood $M \approx
0.3\msun$, $n \approx 0.03 {\rm pc}^{-3}$ and $\sigma_v \approx
50$~ \kms. For the field therefore, $\dot{N}_B \approx 4 \times
10^{-21}\,{\rm pc}^{-3}\,{\rm Gyr}^{-1}$. This shows that the
formation of binaries in the field is extremely rare. Note that in
dense star clusters the stellar density is high and the velocity
dispersion modest, such that the number of binaries formed via
three-body interactions (Eq.~\ref{equation:hut}) may be
substantial. However, {\em wide} binaries, which have semi-major axes
comparable to the size of these clusters, are not formed, as they
simply do not fit in these star clusters. Furthermore, $N$-body
simulations by \cite{kroupaburkert2001} have shown that the
observed broad period distribution of binaries in the field cannot be
produced by dynamically modifying a tighter period distribution in a
star cluster.

A third possibility for the origin of wide binaries is formation
during cluster dissolution, which is the mechanism we propose in this
paper. In an evolving star cluster, stars that are initially unbound,
may become bound to each other as the cluster expands, i.e., if the
gravitational influence of the other cluster members
decreases\footnote{Interestingly, Levison et al. (2009) have
  independently proposed that large populations of comets may be
  captured by stars during cluster dissolution, by a mechanism which
  is similar to that discussed in this paper for wide binary formation
  \citep[see also][]{eggers1997}.}. In order to form a binary pair in
this way, (i) the two stars need to be sufficiently close together,
(ii) the two stars need to have a sufficiently small velocity
difference, and (iii) the newly formed binary should not be destroyed
by gravitational interaction with the remaining cluster stars or field
stars.

Throughout this paper we refer to binaries with $10^3~\mbox{au} < a <
0.1$~pc as the {\em wide binary population}. The paper is organised as
follows. In \S~\ref{section:howmany} we briefly discuss surveys of
wide binary systems and the corresponding observational techniques. In
\S~\ref{section:method} we explain our technique and assumptions. In
\S~\ref{section:analytic} we provide analytical and Monte Carlo
estimates for the resulting wide binary population, and in
\S~\ref{section:nbodysimulations} we present estimates based on
$N$-body simulations of evolving star clusters. Finally, in
\S~\ref{section:conclusions} we present and discuss our conclusions.


\section{Observations of wide binaries}\label{section:howmany}


\begin{figure}
  \centering
  \includegraphics[width=0.5\textwidth,height=!]{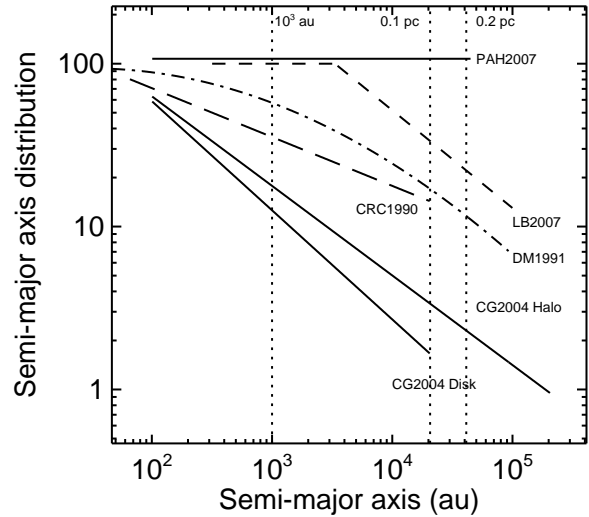}
  \caption{The observed semi-major axis distribution, $f(a)$, of the
  wide binary population (in arbitrary units). The curves indicate the
  \protect\cite{duquennoy1991} distribution (assuming a mass of
  $1~\msun$ for each binary), and the results from
  \protect\cite{close1990}, \protect\cite{lepine2007} (for both the
  Galactic disk and halo), \protect\cite{chaname2004}, and
  \protect\cite{poveda2007}.
  \label{figure:wide_obs} }
\end{figure}

\begin{figure}
  \centering
  \includegraphics[width=0.5\textwidth,height=!]{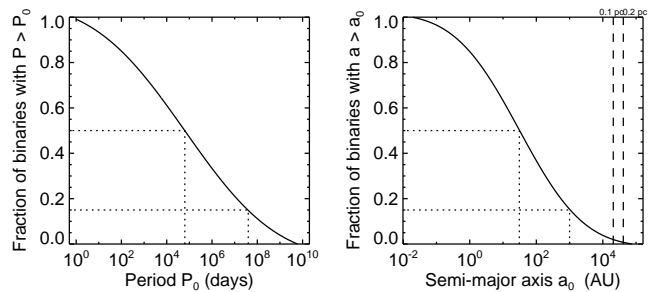}
  \caption{The cumulative distributions in period and semi-major axis for
  solar-type stars in the solar neighbourhood, for the
  \protect\cite{duquennoy1991} log-normal period distribution. The
  total mass of each binary is assumed to be $M_T=1\msun$. The dotted
  lines indicate the median semi-major axis, and $a=10^3$~au.
  \label{figure:cumulative} }
\end{figure}

Observations have indicated that binaries as wide as 1~pc exist in the
halo, while in the Galactic disc the widest binaries have separations
of order 0.1~pc \citep[e.g.,][]{close1990, chaname2004}. Wider
binaries are rare, although some authors claim evidence for binary and
higher-order multiple systems wider than 0.1~pc
\citep[e.g.,][]{scholz2008,caballero2009,mamajek2009}.

Statistical properties of a wide binary population are often recovered
using the angular two-point correlation function
\citep[e.g.,][]{bahcall1981,
garnavich1988, gould1995, longhitano2010}. The most prominent disadvantages of this
method are the inability to identity individual wide binaries and the
need for a good model for the stellar population that is studied.

Individual wide binary candidates are often identified by their common
proper motion on the sky \citep[e.g.,][and numerous
others]{wasserman1991, chaname2004, lepine2007, makarov2008}. Many
wide binaries were found by Hipparcos \citep{esa1997} as well; see
also \citet{soderhjelm2007}. Their nature can then be further
constrained by measuring the parallax and radial velocity
\citep[e.g.,][]{latham1984, hartigan1994, quinn2009}.

The most well-known survey for binarity is arguably that of
\cite{duquennoy1991}, who carried out a large binarity study for
solar-type stars in the solar neighbourhood. They found a log-normal
period distribution $f(P)$ with a mean $\langle \log P \rangle = 4.8$
and a standard deviation $\sigma_{\log P}=2.3$, where $P$ is the
orbital period in days, in the range $1 \la P \la 10^{10}$ days. Note
that the latter value roughly corresponds to an orbital period of
30~Myr. In this log-normal period distribution, $\sim 15\%$ of the
binaries have a semi-major axis wider than $10^3$~au (see
Fig.~\ref{figure:cumulative}). 

Several observational studies suggest a semi-major axis distribution
of the form $f(a) \propto a^{-1}$, which corresponds to a flat
distribution in $\log a$, also known as \"{O}pik's law
\citep{opik1924,vanalbada1968,vereshchagin1988,allen2000,poveda2004}. In
particular, \cite{poveda2007} find binaries in the field follow
\"{O}pik's law in the separation range $100 \la a \la 3000$~au, and
suggest that a population of very young binaries follows \"{O}pik's
law up to $a \approx 45\,000$~au (0.2~pc).

Many other authors have also found significant wide binary
populations, notably \cite{close1990}, \cite{lepine2007}, and
\cite{chaname2004}.  We summarise the wide binary separation
distribution from various authors in Fig.~\ref{figure:wide_obs}.

The reliability of the observed properties of wide binary populations
remains an issue. It is difficult to confidently establish whether the
two components of a candidate wide binary are truly bound, or whether
it is merely a chance superposition. Due to the long orbital periods
of wide binaries, up to millions of years, it is impossible to
accurately derive orbital properties and therefore confirm the bound
state of the candidate wide binary.  It should be noted that due to
the lack of detailed orbital information, the quoted separations of
wide binaries are usually the instantaneous angular separation. The
true separations may thus be significantly larger than the observed,
{\em projected} separations. On the other hand, an estimate for the
semi-major axis distribution of an ensemble of binary systems can {\em
  statistically} be obtained from the projected separation
\citep[e.g.,][]{leinert1993}.

On the other hand, it is extremely difficult to identify binaries with
distant and faint stellar or substellar companions, due to confusion
with foreground and background stars
\citep[e.g.,][]{chandrasekhar1944}. The wide binary fraction as
identified in the observational papers may therefore be a {\em lower
  limit}, rather than an {\em upper limit}.

Although the exact form of the semi-major axis distribution for very
wide binaries is not yet known (see Fig.~\ref{figure:wide_obs}), two
features appear to be clear: (i) the binary fraction
in the separation range $10^3~\mbox{au} < a < 0.1$~pc is roughly 15\%,
and (ii) a sharp drop-off in the separation range $0.1-0.2$~pc is
present, likely due to dynamical destruction of the most weakly bound
binary systems.


\subsection{Stability of very wide binaries}

Whether a wide binary in the Galactic field is stable or not, depends
primarily on its semi-major axis $a$. Using a Monte Carlo approach,
\cite{weinberg1987} show that binaries with $a=0.1$~pc are able to
survive in the Galactic disk for $\sim 10$~Gyr, roughly the age of the
Galaxy, but they do not find a sharp cut-off in the semi-major axis
distribution. On the other hand, a sharp drop-off is observed for
binaries at $a \approx 0.2$~pc due to interactions with other stars,
molecular clouds, and the Galactic tidal field
\citep[e.g,.][]{bahcall1981,retterer1982, mallada2001, jiang2009}.
\cite{gould2006} explain the slope-change at $\sim 3000$~au in the
separation distribution of \cite{lepine2007} as the result of
dynamical interactions in the Galactic field. Very low-mass binary
systems are more weakly bound than their solar-mass analogues; their
typical separation is therefore expected to be considerably smaller
\citep[e.g.,][see also \citealt{kraus2009}]{burgasser2003}, although
several very low mass binaries with separations up to $\sim 0.1$~pc
have been detected \citep[e.g.,][and references therein]{radigan2009}.


\section{Method and assumptions}\label{section:method}

\begin{table}
  \begin{tabular}{lllll}
    \hline
    \hline
    Model & Structure & $R$ (pc) & $N$ & $Q$ \\
    \hline
    P1v & Plummer                 & $0.1 \leq R \leq 1$  & $10$      & 1/2 \\
    P1e & Plummer                 & $0.1 \leq R \leq 1$  & $10$      & 3/2 \\
    P2v & Plummer                 & $0.1$                & $10 \leq N \leq 1000$ & 1/2 \\
    P2e & Plummer                 & $0.1$                & $10 \leq N \leq 1000$ & 3/2 \\
    \hline
    F1v & Fractal, $\alpha=1.5$   & $0.1 \leq R \leq 1$  & $10$      & 1/2 \\
    F1e & Fractal, $\alpha=1.5$   & $0.1 \leq R \leq 1$  & $10$      & 3/2 \\
    F2v & Fractal, $\alpha=1.5$   & $0.1$                & $10 \leq N \leq 1000$ & 1/2 \\
    F2e & Fractal, $\alpha=1.5$   & $0.1$                & $10 \leq N \leq 1000$ & 3/2 \\
    \hline
    \hline
  \end{tabular}
  \caption{Properties of the models used in this paper. The quantity
  $R$ describes the virial radii of models P1 and P2, while it
  describes the radius of the sphere that includes the fractal
  structure for models F1 and F2.  \label{table:models} }
\end{table}

Given the apparent difficulty of forming extremely wide binaries in star
clusters, and the even greater difficulty in keeping them bound we
suggest that these extremely wide binaries form during the dissolution
of a cluster into the field.  

Many young star clusters do not survive for more than a few~Myr
\citep{ladalada, fall2005, mengel2005,bastian2005}.  Their rapid
destruction is probably due to the expulsion of the residual gas
left-over after star formation, which dramatically changes the cluster
potential \citep[][and references therein]{goodwinbastian2006,
  goodwin2009}.  Stars that are unbound in a cluster potential may
become bound to each other after dissolution as the local density
decreases, thus forming an extremely wide binary.

In this section we describe the cluster models used in the remainder
of this paper.  We use two approaches to investigate the formation
of wide binaries during cluster dissolution: Monte Carlo
simulations (\S~\ref{section:montecarlo}) and $N$-body simulations
(\S~\ref{section:nbodysimulations}). In \S~\ref{section:modelsetup} we
explain our choices for the models used in our analysis, in
\S~\ref{section:binarity} we define the quantities we use to describe
binarity, and in \S~\ref{section:modelstability} we describe the
algorithms we use to ensure the stability of the newly formed wide
binaries.


\subsection{Model setup} \label{section:modelsetup}

We simulate star clusters using the {\tt STARLAB} package
\citep{ecology4}. We draw $N$ single stars from the \cite{kroupa2001}
mass function, $f_M(M)$, in the mass range $0.1 \leq M \leq 50
\msun$. The lower-limit corresponds to the hydrogen-burning limit. The
upper limit is (somewhat arbitrarily) set to $50\msun$.

We perform simulations with varying $N$,
ranging from small stellar systems (or sub-clumps) with $N=10$ to open
cluster-sized systems ($N=1000$). 
We additionally perform simulations
of clusters with different radii ($0.1-1$~pc), to identify the
relation between the initial cluster size and the properties of the
newly formed wide binary population. 
We study two sets of dynamical
models: Plummer models and substructured (fractal) models, which we
describe below. The properties for the subsets of models are listed in
Table~\ref{table:models}.

(i) {\em Plummer models}. The Plummer sphere is often used in star
cluster simulations. In this model, each star is given a certain
position and velocity according to the Plummer model
\citep{plummer1911} with a certain virial radius $R_{\rm V}$. The models are
assigned virial radii of $R_{\rm V}=0.1-1$~pc, which are typical for young
clusters. Plummer models are isotropic, and the stellar velocities
follow roughly a Maxwellian distribution (Fig.~\ref{figure:velocitydiagram}). 

(ii) {\em Fractal models}. Young star clusters show a significant
fraction of substructure \citep[e.g.,][]{larson1995, elmegreen2000,
testi2000, ladalada, gutermuth2005, allen2007}. We set the fractal
dimension $\alpha$ to $1.5$ (fractal). For comparison, a value
$\alpha=3$ corresponds to a homogeneous sphere with radius $R$. 
Each star is assigned a velocity, as described in
\cite{goodwin2004}, such that nearby stars have similar
velocities. As in the Plummer models, each cluster is assigned
a radius in the range $R=0.1-1$~pc. Note, however, the difference
between the definition of $R$ for the two sets of models.

The virial ratio $Q \equiv -E_K/E_P$ of a star cluster is defined as
the ratio between its kinetic energy $E_K$ and potential energy
$E_P$. Clusters with $Q=1/2$ are in virial equilibrium, and those with
$Q<1/2$ and $Q>1/2$ are contracting and expanding, respectively. We
study both clusters in virial equilibrium ($Q=1/2$), as well as
clusters with $Q=3/2$. The latter value for $Q$ is expected for young
clusters with an effective star forming efficiency of 33\%
\citep{goodwinbastian2006}. We perform the simulations until the
clusters are completely dissolved, which is typically of the order of
$20-50$~Myr, the timescale at which the majority of low-mass star
clusters are destroyed \citep[see, e.g.,][and numerous
others]{tutukov1978,boutloukos2003,bastian2005,fall2005}.


\subsection{Binarity and multiplicity} \label{section:binarity}

At first, we will consider star clusters that initially consist of
single stars only, while later
(\S~\ref{section:nbodysimulations_binarity}) we will also include
primordial binaries. We do not study the evolution of star clusters
with primordial higher-order ($N \geq 3$) multiple systems. However,
these higher-order systems do form in our star cluster simulations. In
this case, the following three useful quantities describing the
multiplicity of a stellar population can be used:
\begin{eqnarray}
  \binfrac & = & (B+T+\dots)/(S+B+T+\dots) \\
  \nssf    & = & (2B+3T+\dots)/(S+2B+3T+\dots)\\
  \csf     & = & (B+2T+\dots)/(S+B+T+\dots) \,
\end{eqnarray}
\citep[see, e.g.,][]{reipurth1993, kouwenhoven_adonis}. Here, $S$,
$B$, and $T$ denote the number of single stars, binaries, and triples
in the system. The quantity $\binfrac$ is the multiplicity fraction
(commonly known as the ``binary fraction''). $\nssf$ is the non-single
star fraction, as $1-\nssf$ is the fraction of stars that are
single. Finally, $\csf$ is the companion star fraction, which
describes the average number of companions per system, where
``system'' can refer to a single star or multiple system. The number
of systems is given by $S+B+T+\dots$, while $S+2B+3T+\dots$ denotes
the total number of individual stars. Clusters that {\em only} contain
single stars and binary systems have $\binfrac = \csf$.


\subsection{Detection and stability of wide binary and multiple systems} \label{section:modelstability}

After each simulation, potential binary and multiple systems are
identified as those pairs with negative energy
\citep[see also][]{parker2009}.
A multiple ($N\geq 3$) system can only survive for a considerable
amount of time if (i) the system is internally stable, and (ii) if the
outer orbit is stable against perturbations and tidal forces in the
Galactic field. To ensure internal stability of each level in the
hierarchy of the multiple system, we impose the Valtonen stability
criterion $a_{\rm out}/a_{\rm in} > Q_{\rm st}$, where $a_{\rm in}$
and $a_{\rm out}$ are the semi-major axes of the inner and outer
orbits. \cite{valtonen2008} find that
\begin{equation}
  Q_{\rm st} \approx 3 \left( 1+\frac{M_{\rm out}}{M_{\rm in}}
  \right)^{2/3} \left( 1-e \right)^{-1/6} \left( \tfrac{7}{4} -
  \tfrac{1}{2}\cos i - \cos^2 i \right)^{1/3} \,,
\end{equation}
where $M_{\rm in}$ is the (total) mass of the inner component, $M_{\rm
out}$ the mass of the outer component, $i$ the relative inclination of
the orbits, and $e$ the eccentricity of the outer orbit. For a typical
system consisting of equal-mass stars, a circular outer orbit ($e=0$)
and a prograde outer orbit $i=0$, the above expression reduces to
$Q_{\rm st} \approx 3.7$. Systems with $a_{\rm out}/a_{\rm in} >
Q_{\rm st}$ are internally stable for at least $10^4$ revolutions of
the outer component. For wide binaries, with orbital periods of $\sim
500\,000$~years ($a_{\rm out} \approx 10^4$~au), this corresponds to
an {\em internally} stable period of at least 2~Gyr. 

Wide orbits may additionally be unstable against the tidal forces in
the Galactic field and interactions with other single stars and
binaries. We therefore additionally impose a maximum semi-major axis
of 0.1~pc on the outermost orbit of a binary or multiple, motivated by
the observed wide binary population (see
\S~\ref{section:howmany}). The stability of wider binaries is
difficult to assess. As several binaries wider than 0.1~pc are known,
our predictions may slightly underestimate the wide binary
fraction. The properties of the wide binary populations described in
this paper therefore pertain to binaries in the separation range
$10^3~\mbox{au} < a < 0.1$~pc. Note that these binary systems fall
well in the category ``extremely wide binaries'' in the
\cite{zinnecker1984} classification of orbital separations.


\section{Analytic and numerical estimates}\label{section:analytic}

Before proceeding to the $N$-body simulations in
\S~\ref{section:nbodysimulations}, it is useful to first obtain some
analytical approximations for the prevalence of wide
binaries that form during cluster dissolution as well as their 
orbital characteristics. To this end, we first
obtain rough estimates using an analytical approach
(\S~\ref{section:maxwellian}), and subsequently using a Monte Carlo
approach (\S~\ref{section:montecarlo}).


\subsection{Binary formation in Maxwellian velocity space} \label{section:maxwellian}

\begin{figure}
  \centering
  \includegraphics[width=0.5\textwidth,height=!]{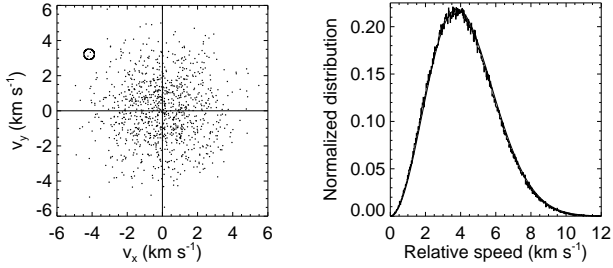}
  \caption{The distribution of velocities and relative speeds for
  Plummer model with $N=1000$, a Kroupa IMF, $Q=1/2$ and a virial
  radius of $R_{\rm V}=0.1$~pc. {\em Left:} The cluster members in the
  $(v_x,v_y)$-diagram, where, as an example, the encircled pair of stars represent a
  potential binary system. {\em Right:} the distribution of relative
  speeds (high-resolution histogram) closely follows the
  Maxwellian distribution (solid curve).
  \label{figure:velocitydiagram} }
\end{figure}

\begin{figure}
  \centering
  \includegraphics[width=0.4\textwidth,height=!]{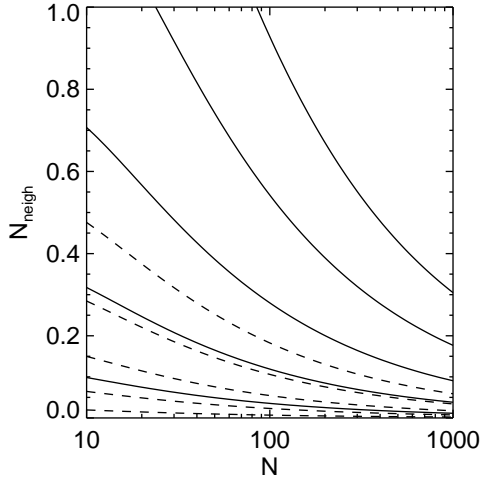}
  \caption{The number of close neighbours to each star (corresponding
  to the number of (potential) binaries formed {\em per star} when
  $N_{\rm neigh} \ll 1$) as a function of $N$, $v_{\rm crit}$ for
  Plummer models with $R_{\rm V}=0.1$~pc. Results are shown for models with
  $Q=1/2$ (solid curves) and $Q=3/2$ (dashed curves). From bottom to
  top, both sets of curves represent the results for $v_{\rm crit} =
  0.10$, 0.15, 0.20, 0.25, and 0.30~\kms.
  \label{figure:maxwell} }
\end{figure}


Wide binaries may form during the dissociation of a cluster if the
relative velocity between two stars is sufficiently small that they
become bound once the perturbing cluster potential is removed. Here we
assume that a wide binary will form if 
the relative velocity is smaller than or roughly
equal to the orbital velocity of a star in a wide binary system. For two
stars with masses $M_1$ and $M_2$ in a circular binary orbit with
semi-major axis $a$, the velocities of the individual stars are given
by
\begin{equation}
v_1 = M_2 \sqrt{\frac{G}{a(M_1+M_2)}} \quad \mbox{and} \quad v_2 =
v_1q^{-1} \,,
\end{equation}
where $G$ is the gravitational constant and $q\equiv M_2/M_1$ is the
mass ratio of the binary system. When adopting, for simplicity,
$q=1$, the above expression reduces to:
\begin{equation}
v_{\rm orb} \approx 30\ \mbox{\kms} \ \left( \frac{M_1+M_2}{\msun} \right)^{-1/2}\left( \frac{a}{\mbox{au}} \right)^{-1/2} \,,
\end{equation}
where $v_{\rm orb}$ is the velocity of either of the components. In
order to be able to form a binary with semi-major axis $a$, we require
velocity differences to be smaller than the critical velocity
\begin{equation} \label{equation:vcrit}
v \la 2v_{\rm orb} \equiv v_{\rm crit}
\end{equation}
For our choice of the IMF, the total mass of a binary system is of
order $1\msun$. Binaries with $a=3.6\times 10^5$, $9\times 10^4$ and
$4\times 10^4$~au
thus typically require velocity differences of $v \la v_{\rm
crit}=0.1$, $0.2$ and $0.3$~\kms, respectively.

If the velocity distribution of the stars in a given star cluster
follows a Maxwell-Boltzmann distribution, then so does also the
distribution of relative speeds between the stars. We define the
relative velocity, $\mathbfss{V}=\mathbfss{v}_i-\mathbfss{v}_j$,
with components $(V_x,V_y,V_z)$ and magnitude $V=|\mathbfss{V}|$. 
The distribution over relative speeds is then given by:
\begin{equation} \label{equation:maxwell}
f_V(V)\,dV = \frac{1}{2 \sqrt{\pi} \sigma^3} \exp \left( - \frac{V^2}{4
\sigma^2} \right) V^2 dV
\end{equation}
\citep[][p.~485]{binneytremaine}, where $\sigma$ is the one-dimensional
velocity dispersion. In the Plummer model $\sigma$ is given by:
\begin{equation} \label{equation:qhalf}
\sigma^2_{Q=1/2} (r) = \frac{16 G \mcl}{18 \pi R_{\rm V}} \left( 1 +
\left( \frac{16r}{3\pi R_{\rm V}} \right)^2 \right)^{-1/2}
\end{equation}
\citep{heggiehut} for a cluster in virial equilibrium (i.e. $Q=1/2$).
Here, $\mcl$ is the total mass of the cluster, and $r$ the distance to
the cluster centre. 
We can re-write Eq.~(\ref{equation:qhalf}) in
units more suitable for the clusters considered in this paper. First,
we set $\mcl=N\langle m \rangle$, where $N$ is the number of stars in
the cluster and $\langle m \rangle$ is their average mass. Using the
\cite{kroupa2001} IMF, $\langle m \rangle = 0.55\msun$. Evaluating the
velocity dispersion at the (intrinsic) half-mass radius, $r =
\halfmass \approx 0.769 R_{\rm V}$ of the cluster, we find:
\begin{equation} \label{equation:sigma_05}
\sigma_{Q=1/2} = 0.64 \times { \left( \frac{0.1 \, \rm{pc} }{ R_{\rm
V}} \right)^{1/2} } { \left( \frac {N}{100 \, {\rm
stars}}\right)^{1/2} } \, \mbox{\kms} \,.
\end{equation}
The kinetic energy for a star cluster with $Q=3/2$ is three times that
of a cluster with $Q=1/2$, and therefore the corresponding velocity
dispersion is
\begin{equation} \label{equation:sigma_15}
\sigma_{Q=3/2} =\sqrt {3} \, \sigma_{Q=1/2} \,.
\end{equation}
As an example, Fig.~\ref{figure:velocitydiagram} shows the distribution
of velocities $(v_x,v_y)$ for a Plummer 
model with $N=1000$ stars, a
virial radius $R_{\rm V} = 0.1$~pc, and $Q=1/2$. The distribution of
relative speeds between random pairs of stars in the cluster is
shown in the right-hand panel. The latter distribution is well
approximated by Eq.~(\ref{equation:maxwell}) with $\sigma=1.9$~\kms,
the velocity dispersion at the half-mass radius is given by
Eq.~(\ref{equation:sigma_05}).

To find the relative fraction $F_b$ of pairs in a given star cluster
which has a relative speed such that they may become bound when the
cluster disperses, we integrate Eq.~(\ref{equation:maxwell})
between $v=0$ and $v=v_{\rm crit}$, where $v_{\rm crit}$ is the
critical velocity difference (Eq.~\ref{equation:vcrit}), below which
we assume that two stars may become bound after cluster dissolution:
\begin{equation}
F_b= \frac{ \int_{0}^{v_{\rm crit}} P(V) dV }{ \int_{0}^{\infty} P(V)
dV} = \frac{\int_{0}^{v_{\rm crit}} \exp \left( - \frac{V^2}{4
\sigma^2} \right) V^2 dV} {\int_{0}^{\infty} \exp \left( -
\frac{V^2}{4 \sigma^2} \right) V^2 dV} \,,
\end{equation}
where we normalised the fraction to unity by dividing by the integral
of Eq.~(\ref{equation:maxwell}) between 0 and $\infty$. To find the
number of pairs with relative speed less than $v_{\rm crit}$ we
multiply $F_b$ by $N-1$. Hence:
\begin{equation}
N_{\rm neigh} = (N-1) \, F_b \,.
\end{equation}
If $N_{\rm neigh}$ is smaller than unity one might expect that the
binary fraction is proportional to $N_{\rm neigh}$. In
situations where $N_{\rm neigh}$ is larger (i.e., larger than unity)
and hence many stars are close to each other in velocity space, we
might expect to have some competition between the stars to stay bound.

As an example we show in Fig.~\ref{figure:maxwell} how $N_{\rm neigh}$
varies with the number of stars, $N$, in clusters with $Q=1/2$ and
$Q=3/2$. We show results in the velocity range $v_{\rm
  crit}=0.1-0.3$~\kms. Depending on their masses and mass ratios,
these velocities correspond to binary systems with semi-major axes
from $\sim 0.1$~pc down to $\sim 10^4$~au.  
Velocities of $1.9$~\kms{}
(not shown in Fig.~\ref{figure:maxwell}) roughly correspond to binary
systems with $a=10^3$~au.  

The values of $N_{\rm neigh}$ in Fig.~\ref{figure:maxwell} are rather
high, as compared to {\em wide} binary fractions derived in
\S~\ref{section:montecarlo} and~\ref{section:nbodysimulations}, mainly
because $N_{\rm neigh}$ also contains companion stars outside the
separation range $10^3~\mbox{au}-0.1$~pc considered throughout this
paper. 
In addition, we believe that the predicted values will drop
further due to the inefficiency of the process. For example, it is not
likely (but also not impossible) that two stars with nearly the same
velocity will form a wide binary system, if there are other stars in
between them. Furthermore, we have only considered the relative
velocities, while we have ignored the relative positions between the
stars. However, this analysis does provide a strong upper limit on the
(wide) binary fractions we might expect after cluster dissolution. 


\begin{figure}
  \centering
  \begin{tabular}{c}
    \includegraphics[width=0.5\textwidth,height=!]{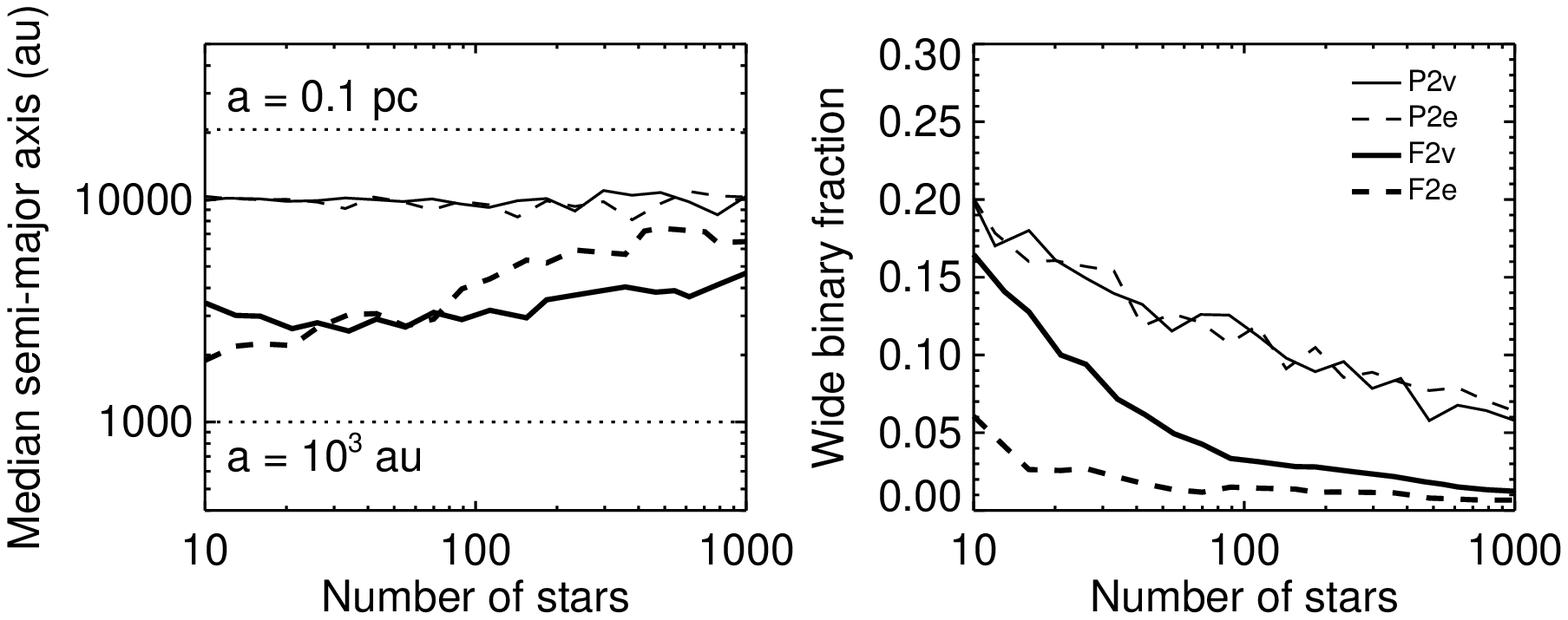} \\
    \includegraphics[width=0.5\textwidth,height=!]{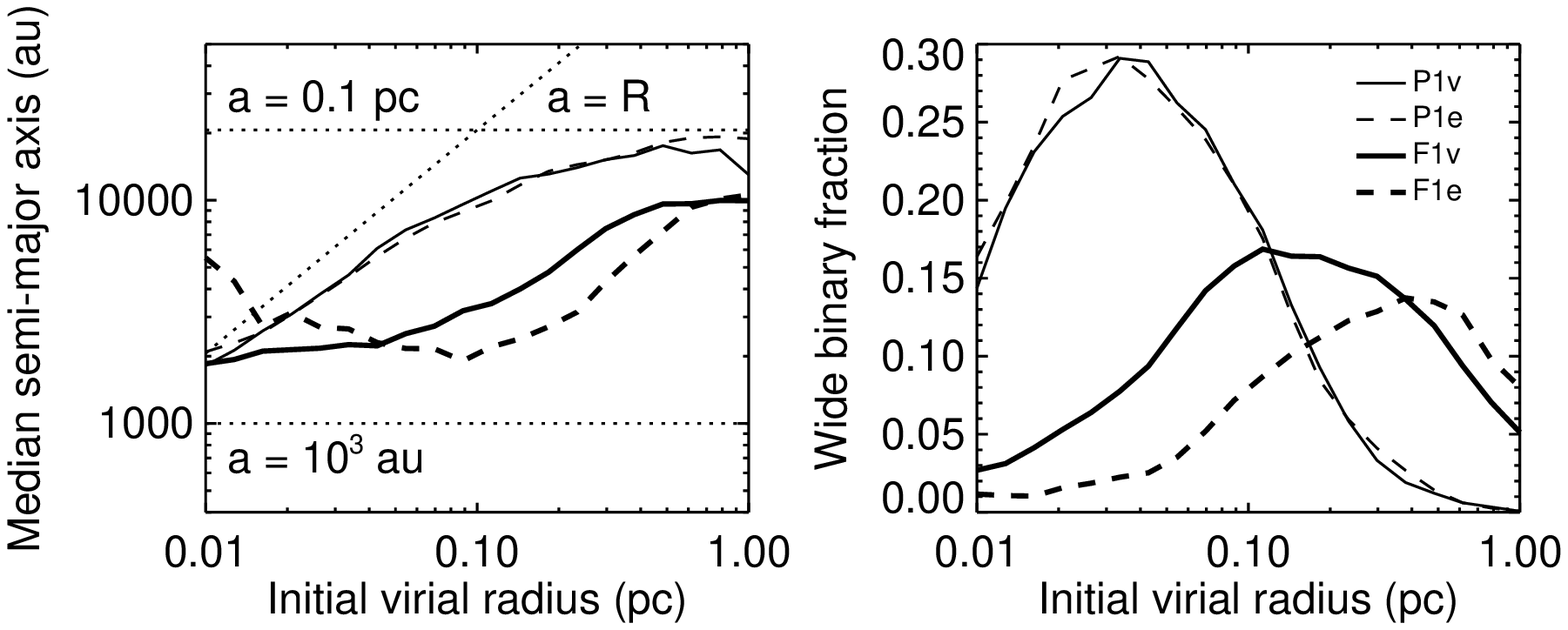}
  \end{tabular}
  \caption{Monte-Carlo predictions for the dependence of the median
  semi-major axis $a$ of the escaping {\em wide} ($10^3~\mbox{au} < a
  < 0.1$~pc) binaries ({\em left}) and {\em wide} binary fraction
  $\binfrac$ ({\em right}) on the number of stars in a cluster ({\em
  top}) and its initial radius $R$ ({\em bottom}).  Results are shown
  for $Q=1/2$ (solid curves) and $Q=3/2$ (dashed curves). The thin and
  thick curves correspond to the Plummer and fractal models,
  respectively. The diagonal dotted line indicates $a=R$. The other
  properties of the modelled clusters are listed in
  Table~\ref{table:models}.
  \label{figure:nr_montecarlo} }
\end{figure}

\begin{figure}
  \centering
  \begin{tabular}{c}
    \includegraphics[width=0.5\textwidth,height=!]{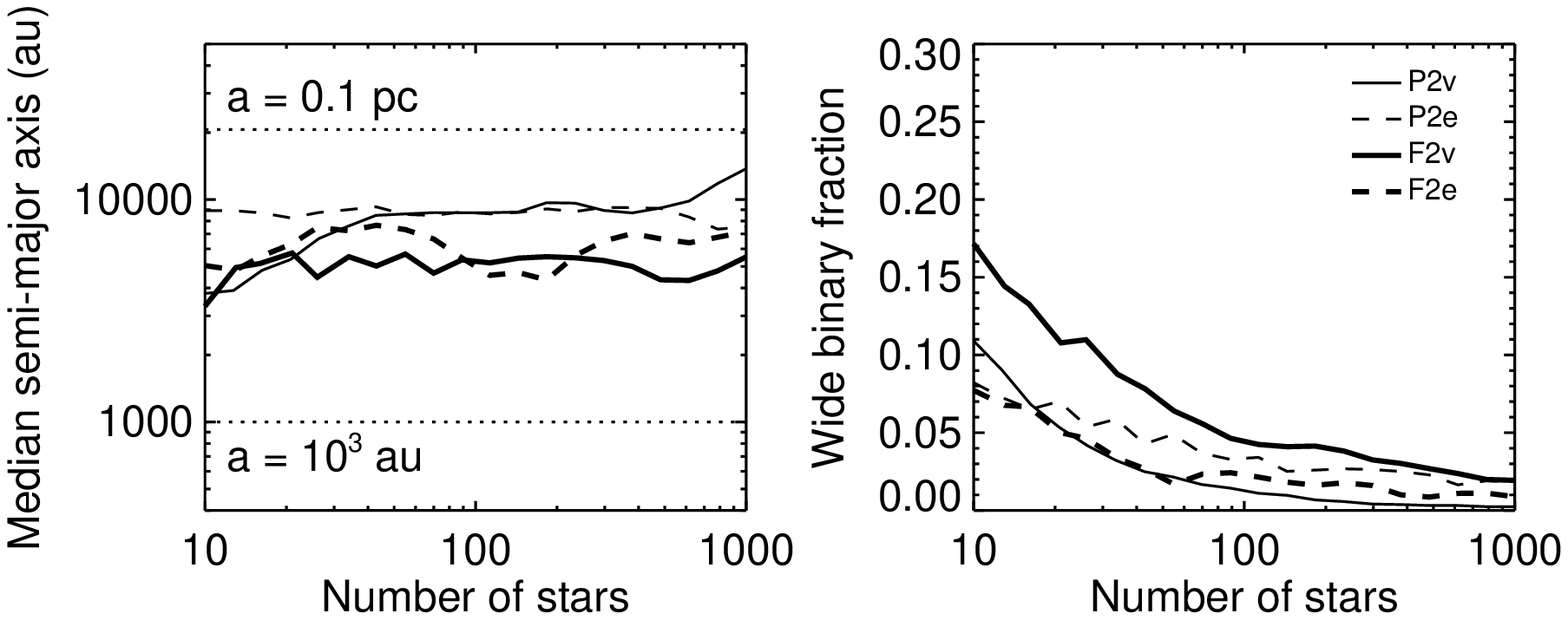} \\
    \includegraphics[width=0.5\textwidth,height=!]{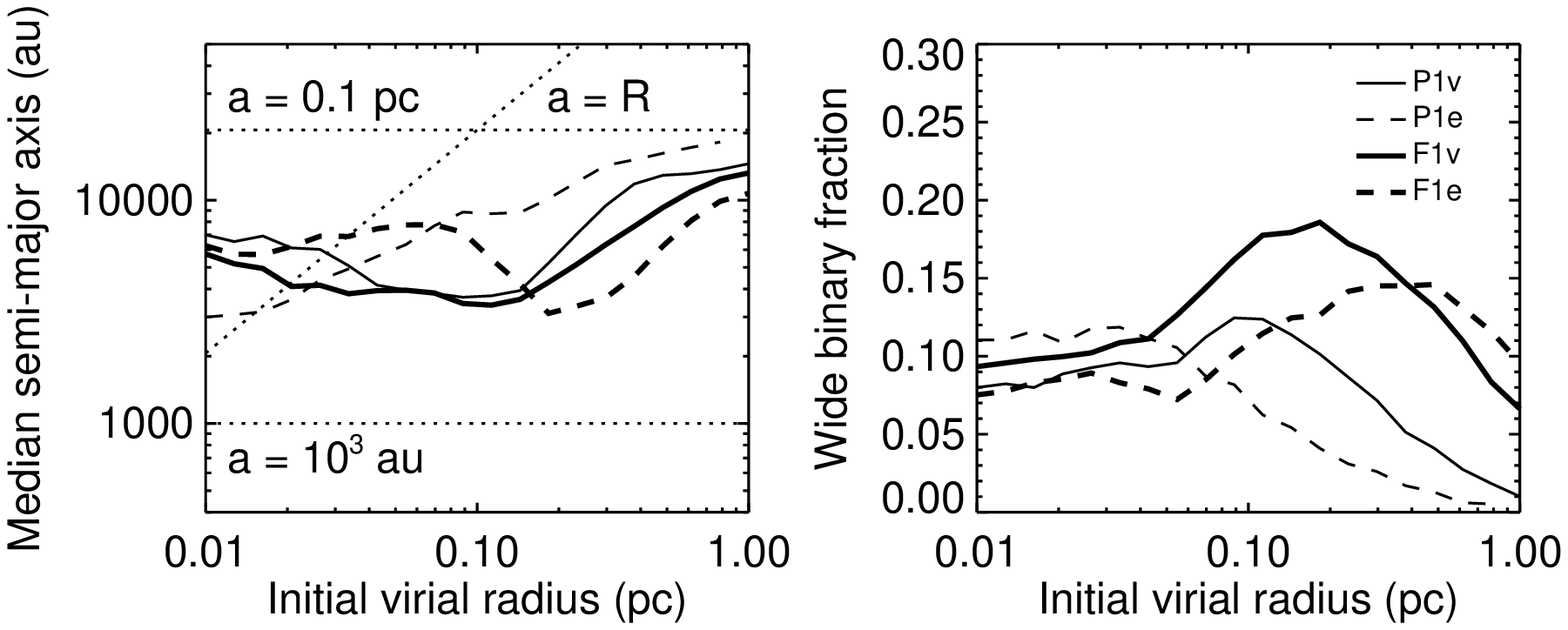}
  \end{tabular}
  \caption{Same as Fig.~\ref{figure:nr_montecarlo}, but showing the
  results of the $N$-body simulations.
  \label{figure:nr_nbody} }
\end{figure}


\subsection{Upper limits from a Monte-Carlo approach} \label{section:montecarlo}

In the previous section we obtained rough estimates for the number of
newly formed binaries using a Maxwellian velocity
distribution. However, we were unable to recover the distributions of
orbital properties, such as the semi-major axis distribution, and we
were not able to take into account the mass spectrum of stars in the
cluster. We therefore use a somewhat more sophisticated Monte Carlo
approach, to obtain estimates for these properties as a function of
cluster size, structure, number of stars, and virial ratio, for the
models listed in Table~\ref{table:models}. Estimates of these
properties are obtained using an ensemble of initial condition
snapshots of each model.

We identify the potential binaries in each star cluster as follows. For
each star with mass $M_1$ we determine its nearest neighbour. More
precisely, we determine the ``most bound'' neighbour, i.e., the
neighbour with mass $M_2$ for which the internal binding energy
\begin{equation}
  E_{\rm b} = \frac{1}{2}\frac{M_1 M_2}{M_1+M_2} v^2 - \frac{G M_1
  M_2}{r}
\end{equation}
is most negative. Here, $r$ is the distance between the two stars, $v$ their
velocity difference, and $G$ the gravitational constant. Subsequently,
we select those pairs of stars that are each other's mutual nearest
neighbours, and assume that they will form a binary system with a
semi-major axis of approximately $r$. Note that not all of these bound
pairs may actually form a binary system, as their velocities are
perturbed by neighbouring stars. We also
ignore the possible presence of triple systems and higher-order systems that
may form. The results, shown in Fig.~\ref{figure:nr_montecarlo},
should therefore be considered as a first-order approximation. We will
discuss this figure in detail in \S~\ref{section:nbodysimulations},
where we will compare the results with those of $N$-body simulations
(shown in Fig.~\ref{figure:nr_nbody}).

Based on a simple Monte Carlo approach, we find that the wide binary
fraction decreases with increasing
stellar density, and mildly decreases with increasing virial
ratio. However, several simplifications have been made, and therefore
these results have to be interpreted with care. In particular, we have
ignored the interaction of each star with all other stars; we have
ignored two-body interactions as well as the tidal field of the
cluster. In the following section we perform a more accurate analysis
to obtain the abundance and properties of wide binaries formed during
cluster dissolution, by performing $N$-body simulations. We will
discuss all properties in detail, and compare these to the results
obtained using the analytical and Monte-Carlo approaches.


\section{Results from $N$-body simulations} \label{section:nbodysimulations}

The previous two sections have shown that wide binary formation during
cluster dissolution may well be possible.  In particular, small dense
clusters seem the most likely to form wide binaries.

In this section we use $N$-body simulations of evolving star clusters
to study how the properties of the newly formed binaries and multiple
systems depend on the initial properties of the clusters. In
\S~\ref{section:nbodysimulations_orbital} we describe the orbital
properties and multiplicity fractions of wide binaries resulting from
a typical star cluster. In \S~\ref{section:nbodysimulations_nr} we
show how the results depend on the initial size $R$ and the number of
stars $N$ in clusters consisting of initially single stars, for
the models listed in Table~\ref{table:models}. In
\S~\ref{section:nbodysimulations_binarity} we study how the
results are affected by the presence of primordial binaries.


\subsection{Properties of the newly formed binary population} \label{section:nbodysimulations_orbital}

\begin{figure*}
  \centering
  \begin{tabular}{c}
    \includegraphics[width=1\textwidth,height=!]{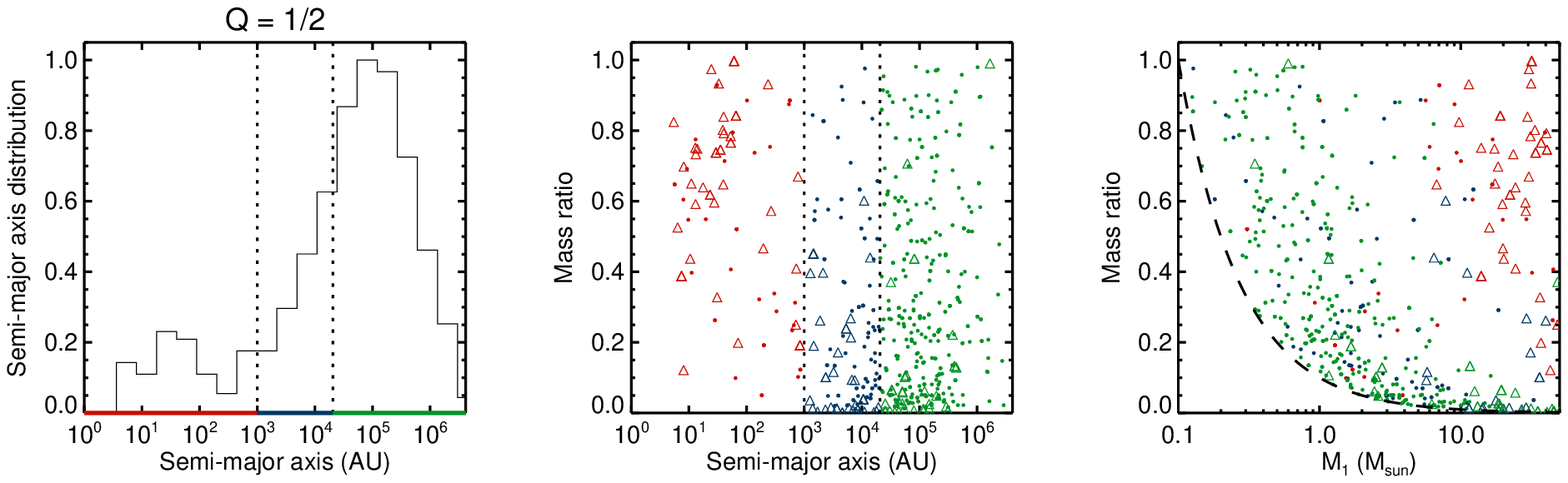}\\
    \includegraphics[width=1\textwidth,height=!]{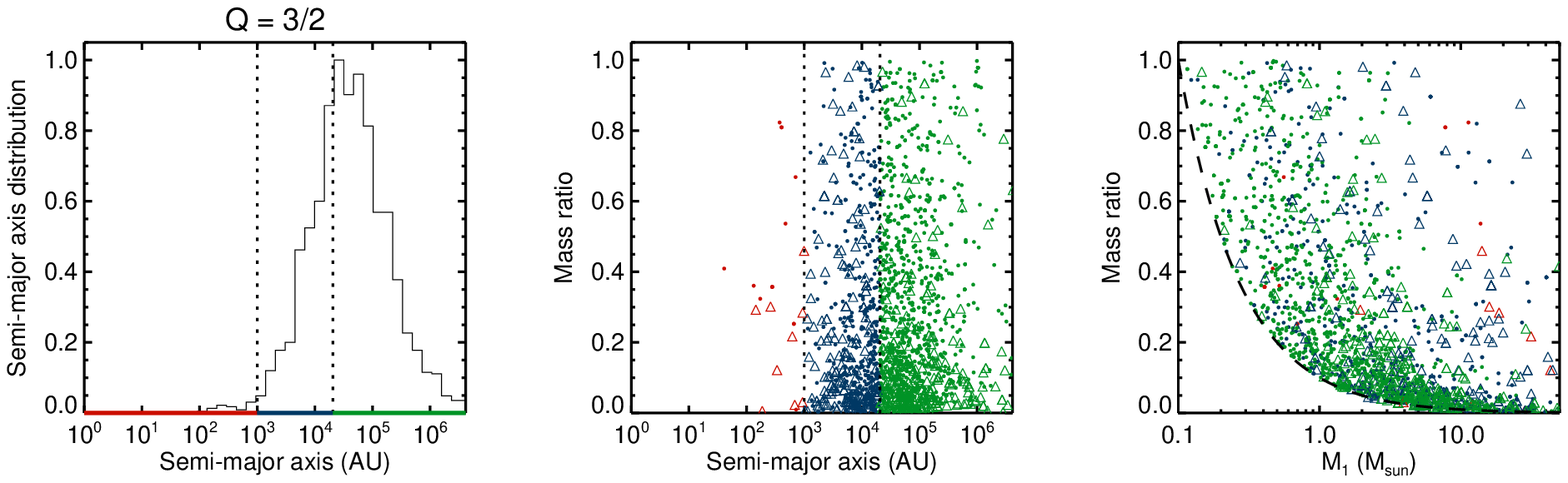}\\
  \end{tabular}
  \caption{The semi-major axis distribution ({\em left}), the
           correlation between mass ratio $q$ and semi-major axis $a$
           ({\em middle}) and between primary mass and mass ratio
           ({\em right}). The histograms in the semi-major axis
           distribution are
           normalized such that the maximum value equals unity. 
           The properties of the orbits of binary
           systems and higher-order multiple systems are indicated with the dots
           and triangles, respectively. For each multiple system with $n$
           stellar components, we
           have included all $n-1$ orbits. Results are shown for 50
           Plummer models with $N=1000$ and $R=0.1$~pc, and virial 
           ratios of $Q=1/2$
           ({\em top}) and $Q=3/2$ ({\em bottom}). The vertical dashed
           lines indicate $a=10^3$~au and $a=0.1$~pc,
           respectively. The dashed curve in the right-hand panel
           indicates the minimum mass ratio $q_{\rm min}(M_1)= M_{\rm
           min} / M_1$.
         \label{figure:bimodal_plummer} }
\end{figure*}

\begin{figure*}
  \centering
  \begin{tabular}{c}
    \includegraphics[width=1\textwidth,height=!]{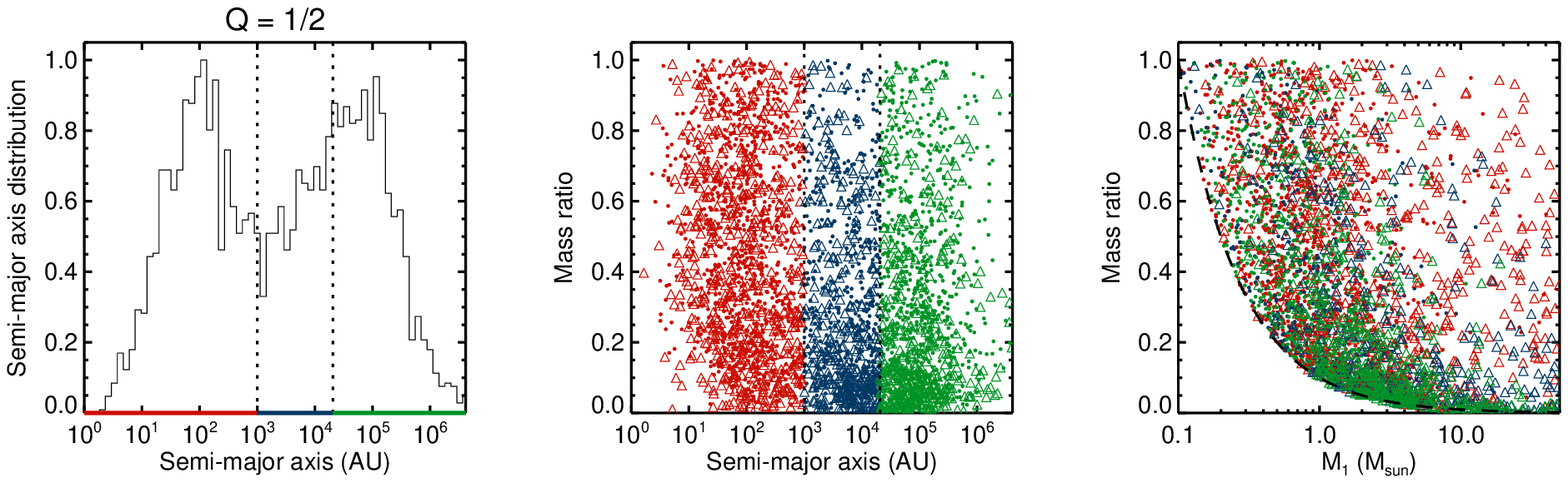}\\
    \includegraphics[width=1\textwidth,height=!]{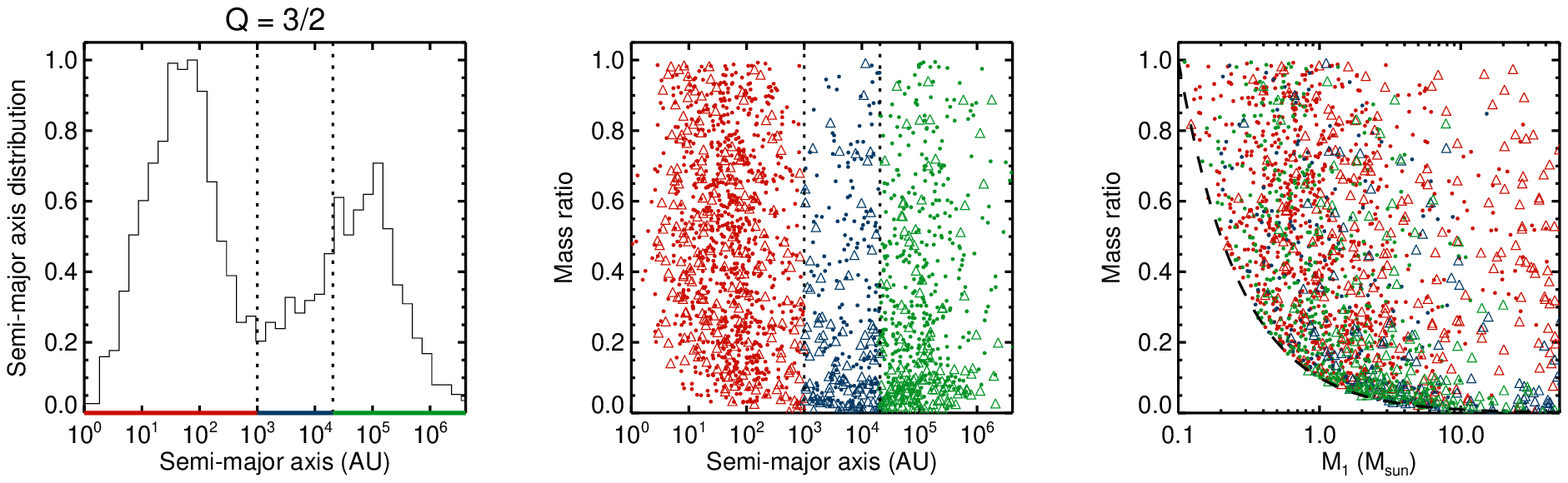}\\
  \end{tabular}
  \caption{Same as Fig.~\ref{figure:bimodal_plummer}, but now for fractal
           models with $N=1000$ and $R=0.1$~pc , and virial ratios 
           of $Q=1/2$ ({\em top}) and $Q=3/2$ ({\em bottom}); in each
           case fifty realisations have been simulated.
    \label{figure:bimodal_fractal} }
\end{figure*}

\begin{figure}
  \centering
  \begin{tabular}{c}
    \includegraphics[width=0.5\textwidth,height=!]{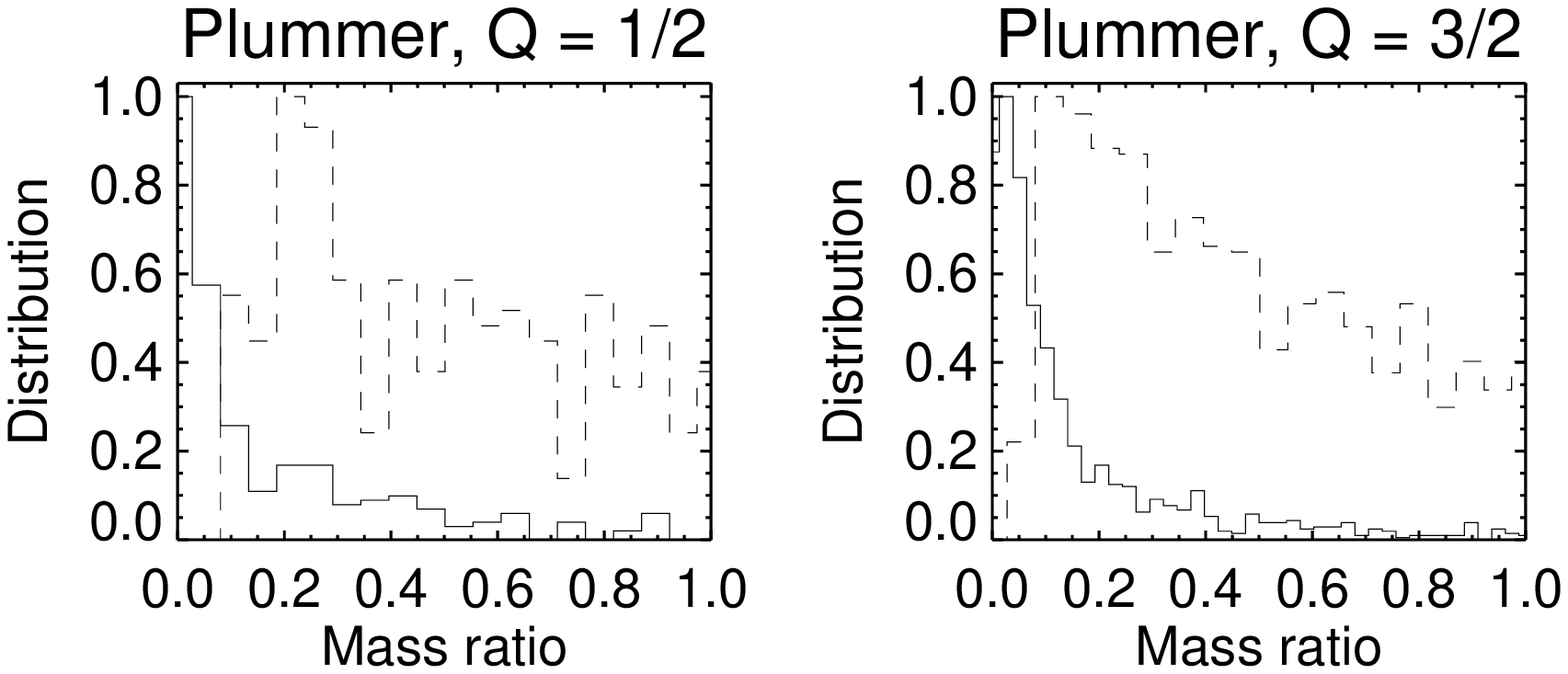} \\
    \includegraphics[width=0.5\textwidth,height=!]{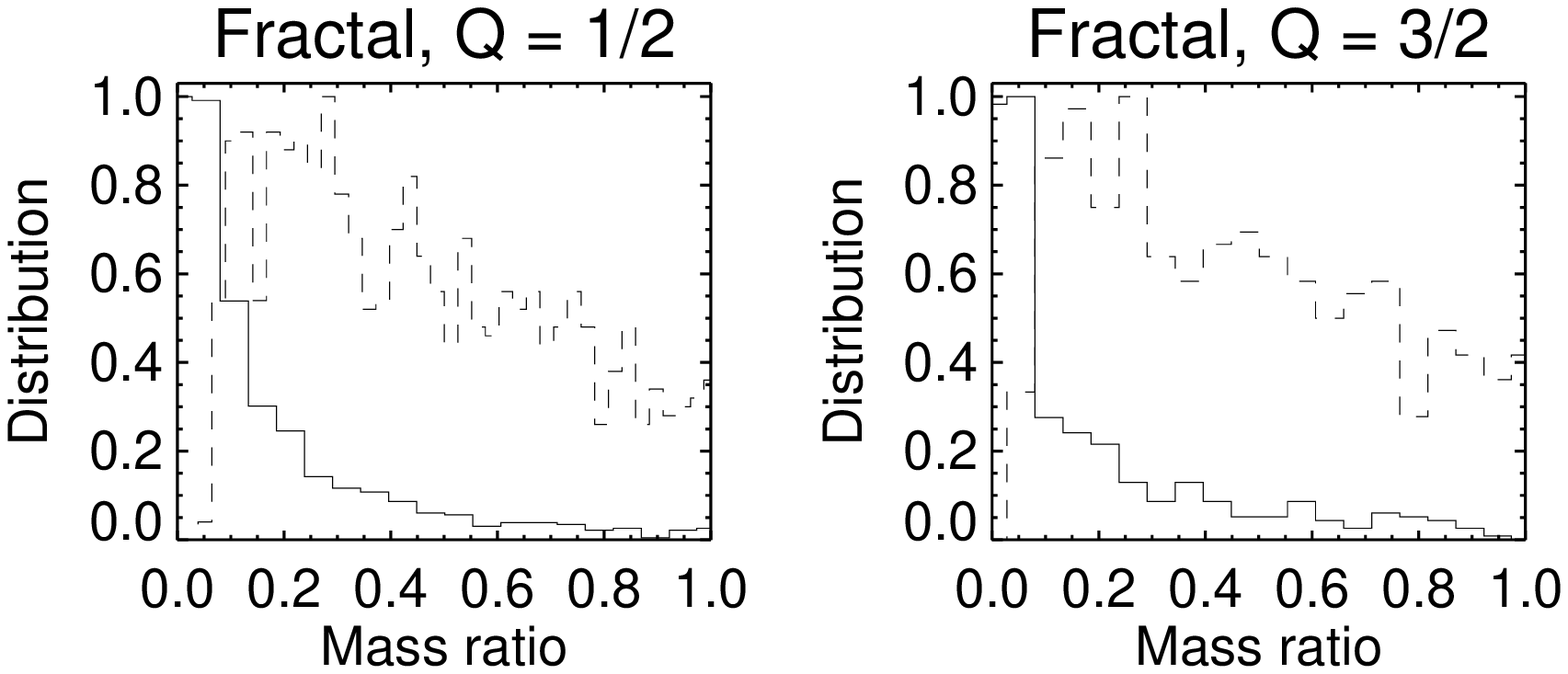} 
  \end{tabular}
  \caption{The mass ratio distribution for binary and multiple systems
           with $a > 10^3$~au resulting from the models in
           Figs.~\ref{figure:bimodal_plummer} and
           ~\ref{figure:bimodal_fractal}.  The solid and dashed curves
           indicate the mass ratio distributions for binaries with
           $M_1>1.5\msun$ and $M_1<1.5\msun$, respectively.
  \label{figure:qqq} }
\end{figure}

\begin{figure}
  \centering
  \begin{tabular}{c}
    \includegraphics[width=0.5\textwidth,height=!]{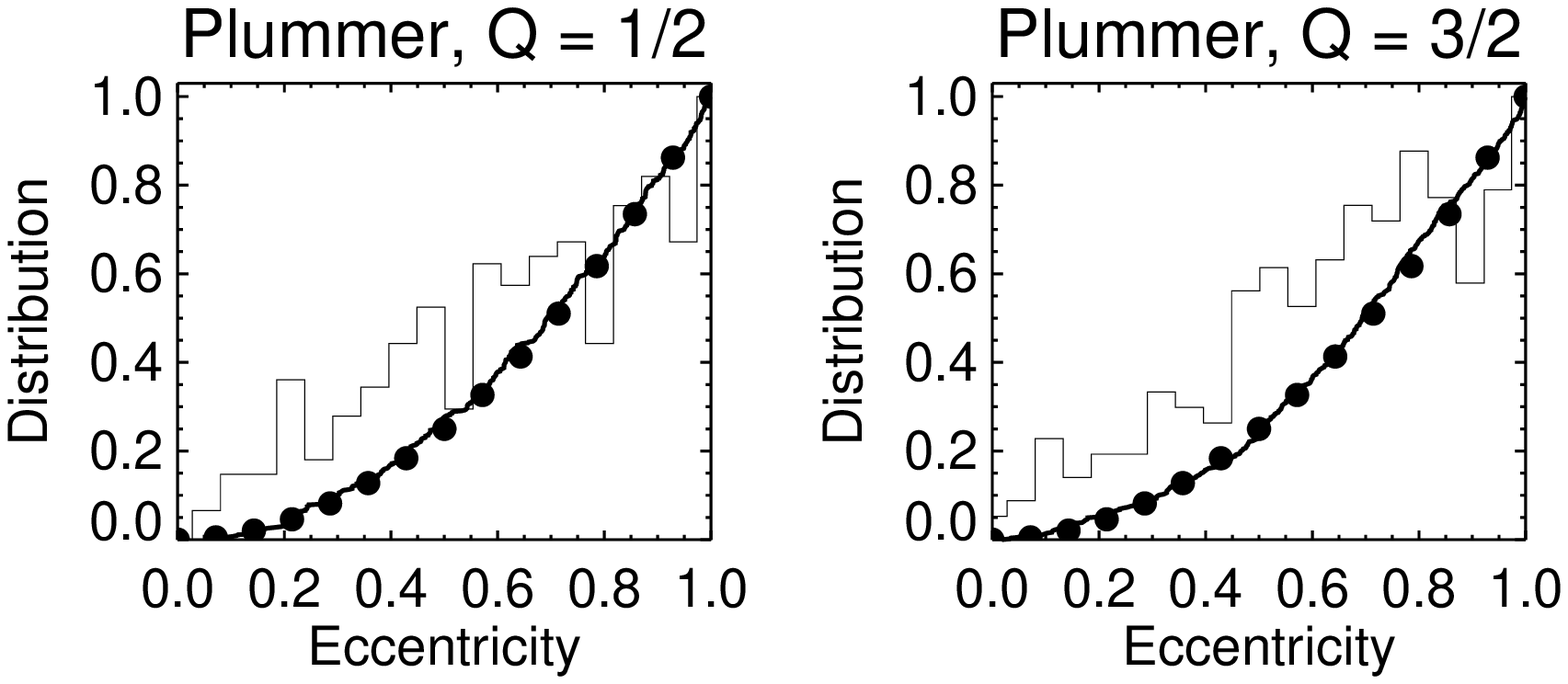} \\
    \includegraphics[width=0.5\textwidth,height=!]{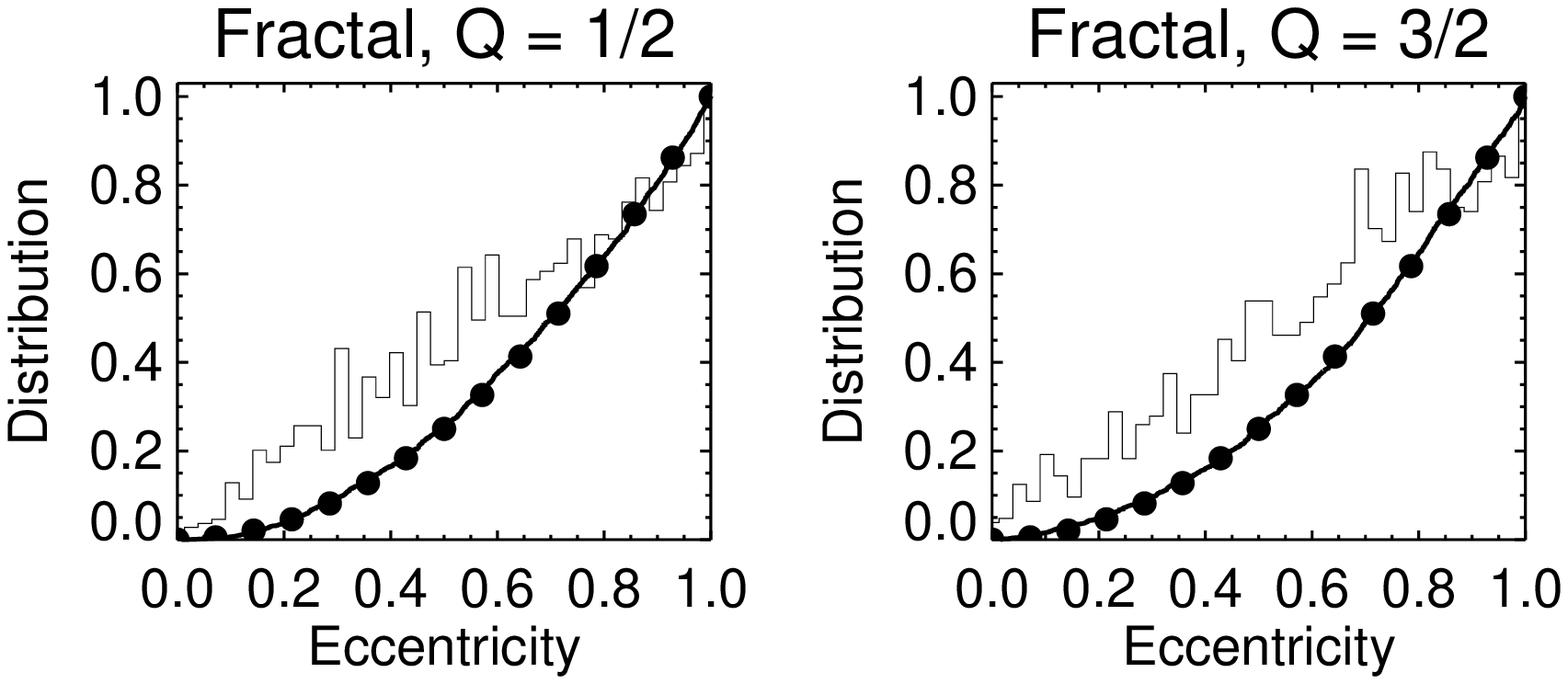} 
  \end{tabular}
  \caption{The eccentricity distribution $f(e)$ for binary and
           multiple systems with $a > 10^3$~au resulting from the models in
           Figs.~\ref{figure:bimodal_plummer} and
           ~\ref{figure:bimodal_fractal}. The histograms indicate
           $f(e)$, while the curves show the corresponding cumulative
           distributions. The filled circles represent the cumulative
           thermal eccentricity distribution $f(e) = 2e$.
  \label{figure:eccentricity} }
\end{figure}


We perform $N$-body simulations of Plummer and fractal clusters consisting of
$N=1000$ stars with radii $R=0.1$~pc (the virial radius for the
Plummer models, or the total radius of fractal models), with initial 
virial ratios $Q=1/2$ and $Q=3/2$ (i.e.,
models P2v, P2e, F2v and F2e in Table~\ref{table:models}). Fifty
realisations of each model are performed to improve the statistics. The
resulting distributions over mass, mass ratio, and semi-major axis for
the resulting binary population are shown in
Figs.~\ref{figure:bimodal_plummer} (Plummer models)
and~\ref{figure:bimodal_fractal} (fractal models).


The left-hand panels in Figs.~\ref{figure:bimodal_plummer} and
~\ref{figure:bimodal_fractal} show the separation distribution $f(a)$
of the resulting binary population. Note that binaries of all
separations are included in these figures, irrespective of whether
they are actually able to survive in the Galactic field or not. The figures
illustrate that the separation distribution of the newly formed
binaries is bimodal, and consists of a small-separation
\dynamicalpeak{} and a large-separation \dissolutionpeak{}. The
tighter binaries in the \dynamicalpeak{} are formed during dynamical
encounters in the cluster, and most of them remain mutually bound
during the further evolution of the cluster.  The wide binaries in the
\dissolutionpeak{}, on the other hand, are formed during the
dissolution phase of star clusters.


The two sets of models with a Plummer density distribution result in a small \dynamicalpeak{},
indicating that dynamical interactions during the lifetime of the
cluster generally do not result in the formation of close
binaries. This is not surprising, as all stars in the Plummer models
are initially given random velocities. Due to the immediate expansion
of the Plummer model with $Q=3/2$, the \dynamicalpeak{} is completely
absent in this case. 

For the models with a fractal density distribution in
Fig.~\ref{figure:bimodal_fractal}, there are numerous binaries in the
dynamical peak. The fact that the \dynamicalpeak{} is stronger for
models F2v/F2e than for models P2v/P2e is due to both the initial 
positions and the initial velocities being correlated in the
fractal models.  Although the average distance between two random
stars is similar for both models, the average distance between {\em
nearest neighbours} in the fractal models is smaller (as they are 
clumpy).   As nearby stars in the clumpy structure also have
similar velocities, frequent dynamical interactions occur, resulting
in a strong \dynamicalpeak{}.


For our choice of initial conditions, binaries in the \dynamicalpeak{}
have separations in the range $1-10^3$~au, with a median value near
$50-100$~au. The median value is set by the typical distance between
stars in the most densely populated regions of the cluster during the formation of
these binary systems. Interestingly, this also corresponds to the
observed peak in the Taurus-Auriga binary separation distribution
\citep[e.g.,][]{leinert1993, kroupa1999}.

Binaries in the \dissolutionpeak{} have a semi-major axis in the
separation range $10^3~\mbox{au}-5$~pc. The widest binaries
in the \dissolutionpeak{} will immediately break up in the Galactic
field, hence our choice to study the wide binary population in the
separation range $10^3~\mbox{au}-0.1$~pc throughout this paper. The
median separation of binaries in the \dissolutionpeak{} occurs at
$a\approx 0.1-0.2$~pc. As we will see later
(\S~\ref{section:dependenceonclustersize}), this value is set by the
initial size of the cluster. 


\begin{table}
  \caption{The specific binary fraction $\binfrac$ for the models
  shown in Figs.~\ref{figure:bimodal_plummer}
  and~\ref{figure:bimodal_fractal}, in which the three ranges in
  semi-major axis are divided with the vertical dotted lines.
  \label{table:bimodel} }
  \begin{tabular}{l ccc}
    \hline
    \hline
    Model &       $\binfrac$         & $\binfrac$                          &   $\binfrac$            \\
    Separation range & $<10^3$~au & $10^3~\mbox{au} - 0.1$~pc & $> 0.1$~pc \\
    \hline
    P2v ($N=1000$) & 0.2\%   & 0.3\%   & 0.8\%   \\
    P2e ($N=1000$) & 0.1\%   & 1.4\%   & 2.8\%   \\
    F2v ($N=1000$) & 3.3\%   & 1.8\%   & 2.6\%   \\
    F2e ($N=1000$) & 2.2\%   & 0.6\%   & 1.1\%   \\
    \hline
    \hline
  \end{tabular}
\end{table}


For practical purposes, we consider three ranges in semi-major axis:
close binaries with $a<10^3$~au, wide binaries with $10^3~\mbox{au}
<a< 0.1$~pc, and extremely wide binaries with $a> 0.1$~pc. The limits
are indicated with the vertical dotted lines in the figures. Most
close binaries that are found in star clusters are formed via the
``normal'' star formation process, with the small number seen in these
simulations formed by dynamical interactions. The
wide and extremely wide binaries are formed during the cluster
dissolution phase. Note however, that the vast majority of the
extremely wide binaries are unstable in the Galactic field, and are
ionised quickly after their formation.


For the models in Figs.~\ref{figure:bimodal_plummer}
and~\ref{figure:bimodal_fractal}, the specific binary fraction (i.e.,
the fraction of binary systems in a certain semi-major axis range) of
the three types of binaries are listed in
Table~\ref{table:bimodel}. The highest wide binary fractions of a few
per cent (in the separation range $10^3~\mbox{au} - 0.1$~pc) are
obtained for Plummer models with $Q=3/2$, and fractal models with $Q=1/2$.


The middle and right-hand panels of Figs.~\ref{figure:bimodal_plummer}
and~\ref{figure:bimodal_fractal} show the correlations between
semi-major axis, mass ratio, and primary mass, for the binary and
multiple (higher-order) systems in each of the models. The panels
indicate the presence of a large number of newly formed multiple
systems. These higher-order systems are stable in isolation, but a
large fraction will not be able to survive in the Galactic field,
where tidal forces will rapidly remove the outer component from the
system. Figs.~\ref{figure:bimodal_plummer}
and~\ref{figure:bimodal_fractal} therefore overestimate the fraction
of higher-order multiple systems. Note, in particular, the high
prevalence of multiple systems in the \dynamicalpeak{}. Many outer
components of these systems fall in the \dissolutionpeak{}. These
systems are therefore wide higher-order systems.


For the Plummer models, the \dynamicalpeak{} consists of systems with
high masses and high mass ratios. This is a well-known signature of
mass segregation: the highest-mass stars sink to the cluster centre,
where they form close binaries \citep[e.g.][]{heggiehut}. During the
dissolution phase of the clusters, these close, massive binaries act
like single stars when forming a ``wide binary'', which is in fact a
wide triple or higher-order multiple system. The effect of mass
segregation is less visible for the fractal models, where dynamical
interactions in the subclumps play a greater role. However,
Fig.~\ref{figure:bimodal_fractal} still clearly shows that most
massive systems are mostly close ($a<10^3$~au) and often higher-order. In
addition to the triple and higher-order systems formed during the
dissolution process, several higher-order systems may form via
dynamical interactions \citep{vandenberk2007}.


The correlations between the primary mass and mass ratio distributions
for the binary and multiple systems are similar to those expected from
random pairing of individual stars. To first order approximation, the
masses of the two stars, $M_1$ and $M_2$, in each binary are
uncorrelated; one would therefore expect something similar to random
pairing \citep[e.g.,][]{kouwenhoven_pairing}, where the average mass
ratio decreases with increasing binary system mass. The resulting mass
ratio distributions for binary and multiple systems with $a>10^3$~au
(i.e., those in the \dissolutionpeak{}) are shown in
Fig.~\ref{figure:qqq}, which illustrates the dependence of the mass
ratio distribution on mass.

Based on the analysis of a sample of 798 common proper motion pairs,
\cite{trimble1987} also come to the conclusion that the very wide
binary population in the field is consistent with random pairing, and
\cite{valtonen1997} come to the same conclusion from their simulations
of three-body encounters. However, the wide binary population does not
result from random pairing alone, as the interaction between two stars
depends on their mutual gravitational attraction, and the probability
of two stars forming a binary is thus proportional to the product
$M_1M_2$. In other words, gravitational focusing
\citep[e.g.,][]{gaburov2008} plays an important role.


Measurements of the eccentricity distribution of wide binaries 
are currently unavailable, due to the large orbital periods
and incompleteness. If we suspect that the vast majority of wide 
binaries probably have formed dynamically, and as dynamical 
interactions are common among the
widest binaries (with respect to closer-in binaries), the best guess
is perhaps the thermal eccentricity distribution $f(e) = 2e$ $(0 \leq
e < 1)$ \citep[][see also \citealt{kroupa2008} for a derivation]{heggie1975}, which results from energy equipartition.
The eccentricity distributions resulting for binaries in the
\dissolutionpeak{} ($a>10^3$~au) are shown in
Fig.~\ref{figure:eccentricity}. As expected, the thermal eccentricity
distribution is a good approximation for the newly formed binary
population.


If wide binaries form during the dissolution process of a star
cluster, then the orbital and spin angular momenta of
the components should be randomly aligned.  On the other hand, if the
two components formed together in some way it might be that the
orbital and spin angular momenta of the components will
be correlated (as seen for example in the observations of $\sim
100$-au Ae/Be binaries by \citealt{baines2006}).  Therefore, observations
of the relative alignments of orbital and spin angular momenta could
provide constraints on the possible formation mechanisms of
very wide binaries.


Finally, the age difference (between primary and companion star) for a
population of wide binaries could provide a clue to their origin
\citep[see, e.g.,][]{kraus2009b}. For a star cluster with a certain
age spread, one might expect the components of the resulting wide
binary population to exhibit a similar age difference. 
This age
difference is measurable, but only for young ($\lesssim
10$~Myr) binary systems. On the other
hand, this age difference may be smaller than expected from random
pairing, if an initial correlation between position and velocity
exists. 


\subsection{Dependence on cluster properties} \label{section:nbodysimulations_nr}

In this section we describe how the properties of the wide binary
population depend on the initial conditions we assign to a star
cluster, in particular its size $R$, number of stars $N$, virial ratio $Q$, and
morphology (Plummer sphere or fractal structure). We adopt the cluster
properties listed in Table~\ref{table:models}. We compare the results
that we derived earlier using Monte Carlo simulations
(Fig.~\ref{figure:nr_montecarlo}), with the results of $N$-body
simulations, shown in Fig.~\ref{figure:nr_nbody}.


\subsubsection{Dependence on the initial cluster mass} \label{section:clustermass}

The top panels of Fig.~\ref{figure:nr_nbody} show the median
semi-major axis $\amed$ and the binary fraction $\binfrac$ {\em of
  wide binaries} ($10^3~\mbox{au}<a<0.1$~pc) as a
function of the number of stars $N$ in a cluster.
For both the Plummer models and the fractal models, $\amed$ does not
vary significantly with $N$ and the virial ratio $Q$. The reason for
this is that all these models have an identical size $R$. 
{\em Since
$R$ is the most important size scale imposed on the modelled star clusters, it
determines the size scaling (i.e., semi-major axis distribution) of
the newly formed binaries.}  

The dependence of $\binfrac$ on $N$ and $Q$ is qualitatively the same
as the analytical predictions shown in Fig.~\ref{figure:maxwell} and
the Monte Carlo approximation shown in Fig.~\ref{figure:nr_montecarlo} . The
wide binary fraction $\binfrac$ decreases with increasing $N$ because
the stars are further apart in velocity space
(cf. Fig.~\ref{figure:velocitydiagram}), i.e., the velocity dispersion
is larger, and hence two neighbouring stars are less likely to form a
bound system. 

For the $N$-body simulations we find that the fractal model with
$Q=1/2$ provides the highest wide binary fractions, although the
difference between models is fairly small (especially when compared to
the difference with increasing $N$). Models with $Q=3/2$
generally result in a smaller wide binary fraction than those with
$Q=1/2$, due to the larger distance between the stars in velocity
space (see Eqs.~\ref{equation:sigma_05}
and~\ref{equation:sigma_15}). The curves for the fractal models in
Figs.~\ref{figure:nr_montecarlo} and~Fig.~\ref{figure:nr_nbody} are
almost the same, indicating that the Monte Carlo approach provides a
good estimate of the wide binary population. For the Plummer models,
the Monte Carlo approach predicts a binary fraction that is too
high, which is due to the fact that the positions and velocities
of stars in the Plummer models as we initialise them are uncorrelated.


\subsubsection{Dependence on the initial cluster size} \label{section:dependenceonclustersize}

The bottom panels of Fig~\ref{figure:nr_nbody} shows the dependence of $\amed$ and
$\binfrac$ on the initial size $R$ of the clusters. Again, these
values are only for {\em wide} binaries with
$10^3~\mbox{au}<a<0.1$~pc. Note the different definitions of $R$: for
the Plummer models $R$ represents the virial radius, while for the
fractal models $R$ represents the radius of the sphere enclosing the
whole system.  Note again the similarity between the Monte Carlo
approximation shown in Fig.~\ref{figure:nr_montecarlo} and the
$N$-body models.

As discussed above, the initial cluster size $R$ determines the length
scale in each model, and therefore the size scaling of the semi-major
axis distribution of the newly formed binaries. For example, changing
the initial size of the clusters shown in
Figs.~\ref{figure:bimodal_plummer} and~\ref{figure:bimodal_fractal}
would simply result in the semi-major axis distribution in the
left-hand panels being shifted to smaller or larger values of $a$. 

This direct dependence of $f(a)$ on $R$ is not seen directly in
Fig.~\ref{figure:nr_montecarlo} because we only show the results for
wide binaries in the separation range $10^3~\mbox{au}<a<0.1$~pc, and
because $f(a)$ is bimodal. However, the $R$-dependent median
semi-major axis and binary fraction can be explained by the
\dynamicalpeak{} and \dissolutionpeak{} shifting through the range
$10^3~\mbox{au}<a<0.1$~pc whilst varying $R$.

The highest $\binfrac$ is found when either the \dynamicalpeak{}, or the
\dissolutionpeak{}, is centred in the separation range
$10^3~\mbox{au}<a<0.1$~pc.  For our choice of the initial conditions,
this peak occurs at $R=0.025$~pc for the Plummer models, when the
\dissolutionpeak{} is centred in the range
$10^3~\mbox{au}-0.1$~pc. The peak in $\binfrac$ occurs at $R\approx
0.15$~pc for the fractal models with $Q=1/2$ and at $R \approx 0.6$~pc for
fractal models with $Q=3/2$, when the \dynamicalpeak{} is centred in
the range $10^3~\mbox{au}-0.1$~pc.

Given our set of initial conditions, compact clusters result in a wide
binary fraction of $8-12\%$, irrespective of virial ratio and
morphology. For more extended clusters, those with a Plummer structure
and those with a higher virial ratio result in a smaller binary
fraction.  The difference between the Plummer and fractal models can be
explained by (i) the difference in the definition of $R$ for the two
sets of models, and (ii) by the different {\em intrinsic} separation
distribution (see the left-hand panels in
Figs.~\ref{figure:bimodal_plummer} and~\ref{figure:bimodal_fractal}).

The cluster size $R$ determines the length scale of the system, and
therefore determines the typical semi-major axis of the newly formed
wide binaries. Other, less important length scales in the system are
the mean distance between two stars, which depends on the parameters $R$,
$N$, and the stellar density distribution (see
\S~\ref{section:clustermass}), as well as the typical semi-major axis of
primordial binary systems (see
\S~\ref{section:nbodysimulations_binarity}). 


\subsection{Effects of primordial binarity}\label{section:nbodysimulations_binarity}

\begin{figure}
  \centering
  \begin{tabular}{c}
    \includegraphics[width=0.45\textwidth,height=!]{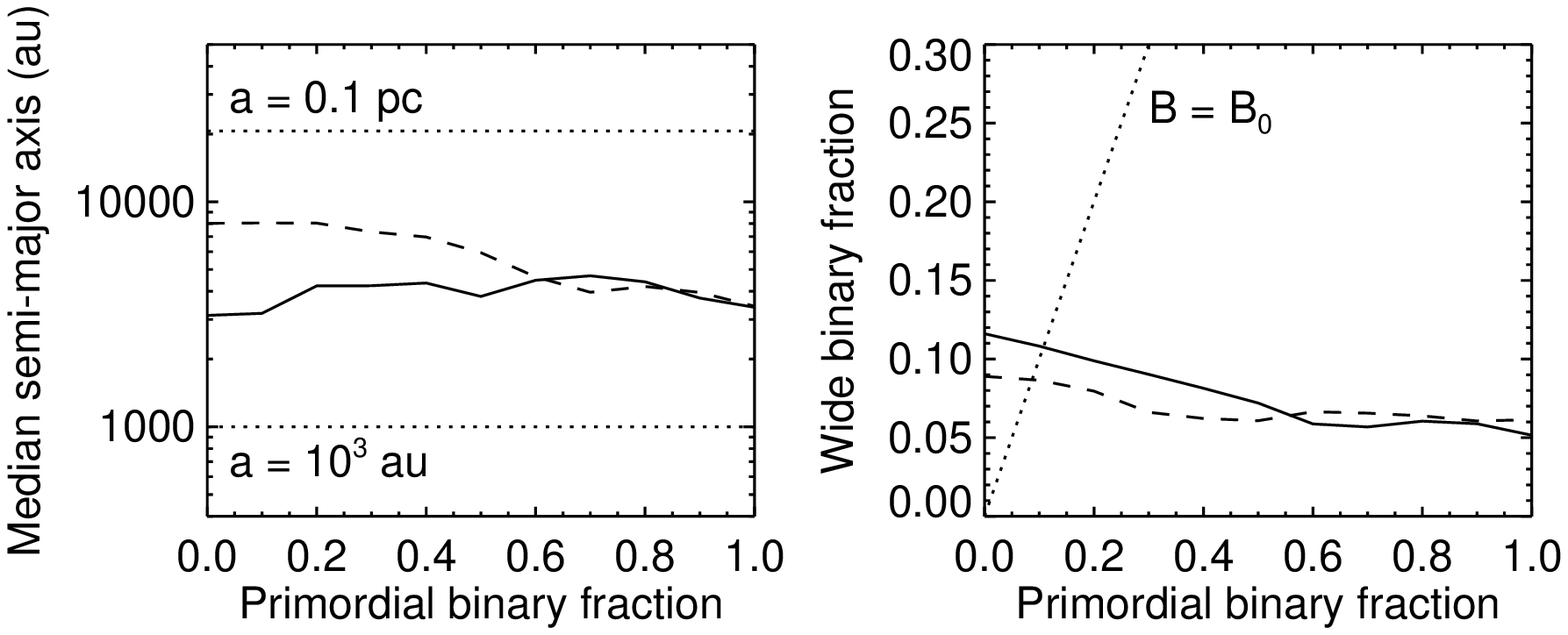} \\
    \includegraphics[width=0.45\textwidth,height=!]{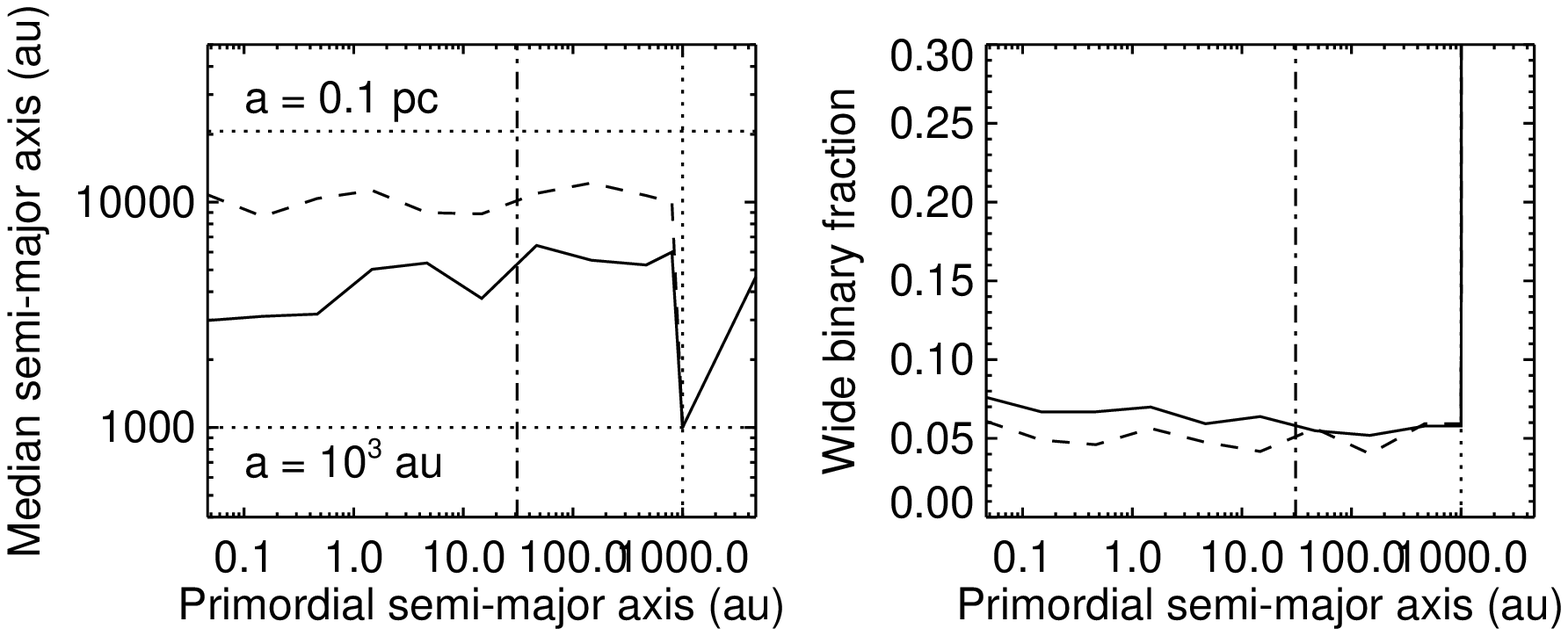}
  \end{tabular}
  \caption{The effect of primordial binarity on the formation of wide
  binaries, for Plummer models with $N=10$ and $R=0.1$~pc. The solid
  and dashed curves in each panel indicate the results for $Q=1/2$ and
  $Q=3/2$, respectively. {\em Top:} the effect of a variable
  primordial binary fraction $\binfrac_0$. The bottom horizontal
  dotted line indicates $a=18.2$~au, the median semi-major axis for
  primordial binaries. {\em Bottom:} the effect of the semi-major axis
  $a_0$ for models with a primordial binary frequency
  $\binfrac_0=50\%$ in which each binary has a semi-major axis
  $a=a_0$. The dash-dotted lines indicate $a=18.2$~au, the median
  semi-major axis of binary systems in the Galactic field. The
  vertical dotted line indicates $a=10^3$~au, beyond which all
  primordial binaries are classified as wide binaries.
  \label{figure:binaritysma} }
\end{figure}


In the analysis above we have considered star clusters that
initially consist of single stars only. The results for star clusters
with a non-zero primordial binary fraction are very similar to the
results described above, with the difference that the components of
the wide ``binary'' are now in many cases primordial binaries. In
other words, the majority of the wide ``binaries'' that formed in the
simulations described in the previous sections, actually describe the
outer orbits of wide triple and quadruple systems.


We predict the properties of these wide triple and quadruple systems
by performing $N$-body simulations of the Plummer models listed in
Table~\ref{table:models}, but now we include a non-zero primordial
binary fraction.
We perform the simulations with primordial binary fractions
$\binfrac_0$ ranging from 0\% to 100\%.  We adopt the
\cite{kroupa1995b} birth period distribution. This distribution is
derived from a detailed analysis of observed stellar populations, and
has the form
\begin{equation}
  f_P(P) = 2.5 \, \frac{
    \left( \log P - \log P_{\rm min} \right)
  }{
    45 + \left( \log P - \log P_{\rm min} \right)^2
  }
\end{equation}
for $P_{\rm min} \leq P \leq P_{\rm max}$, where $\log P_{\rm min} =
1$, $\log P_{\rm max} =8.43$, and $P$ is the period in days. We adopt
a thermal eccentricity distribution $f(e) = 2e$ ($0\leq e < 1$). We
adopt a flat mass ratio distribution $f(q) = 1$ with $0 < q \equiv
M_2/M_1 < 1$ \citep[i.e., we apply pairing function PCP-I;
see][]{kouwenhoven_pairing}. Subsequently, we generate an initial
population from this birth population, by applying eigenevolution as
described in \cite{kroupa1995b}. All binaries are assigned random
orientations and orbital phases at the beginning of the simulations.


Due to the inclusion of binary components, the total mass of each
cluster increases slightly (up to a maximum of $50$\% for a primordial
binary fraction of 100\%), although the number of ``systems'', $N=S+B$
remains constant. Strictly speaking, it is thus not appropriate to
directly compare clusters with and without binaries, as we have
changed more than one parameter: binarity {\em and} cluster mass
\citep[see, e.g.,][]{kouwenhoven_sigma}. However, as the increase in
cluster mass due to adding the companions is rather small, we will
ignore this issue.


The results for clusters with a varying primordial binary
fraction $\binfrac_0$ is shown in the top panels of
Fig.~\ref{figure:binaritysma}. For small binary fractions, the
results are very similar to those of clusters without primordial
binaries.  The properties of the resulting wide binary population
depend mildly on $Q$. An increasing $Q$ results in, on average, wider
binaries, hence in a larger fraction of binaries with $a>0.1$~pc, and
therefore in a slightly smaller wide binary fraction. Note that
$\binfrac$ decreases slightly with increasing $\binfrac_0$. A larger
primordial binary fraction results in a smaller wide binary fraction,
possibly because of the destruction of newly formed wide binaries by
primordial binaries (which have a significantly larger collisional
cross-section than single stars).


Whether or not primordial binarity affects the formation of wide
binaries depends not only on the primordial binary fraction, but also
on the properties of these binaries: the semi-major axis (or period)
distribution, the eccentricity distribution and the mass ratio
distribution. The most important of these is the semi-major axis
distribution $f(a)$, as it determines the internal binding energy of a
binary and the cross-section for gravitational interactions between
binaries and other binaries or single stars.  In order to extract the
dependence on $f(a)$, we simulate clusters in which all binaries have
a single value for $a=a_0$. We vary $a_0$ in each cluster, and
determine the number of newly-formed binaries. In all simulations we
adopt a primordial binary fraction of 50\%, a flat mass ratio
distribution and a thermal eccentricity distribution.


The results of these simulations are shown in the bottom panels of
Fig.~\ref{figure:binaritysma}. For models with $a_0 < 10^3$~au, most
primordial binaries survive, while additional wide binary, triple, and
quadruple systems are formed. In fact, the resulting wide binary
fraction is practically independent of the primordial binary fraction
$\binfrac_0$. For models with $a_0 > 10^3$~au, all primordial binary
systems are classified as wide binaries. For these models we therefore
have $\binfrac \approx \binfrac_0$ and a median semi-major axis equal
to $a_0$, which results in the glitches at $a=10^3$~au in
Fig.~{\ref{figure:binaritysma}.


The wide orbits are part of systems with $2$, $3$, and $4$
components. They are formed by randomly pairing single stars and
primordial binary systems together. The number of multiple systems of
each degree can thus be estimated by simply calculating the
probability of randomly drawing a single-single, single-binary, and
binary-binary pair. When assuming a primordial binary fraction
$\binfrac_0$, the multiplicity distribution of the resulting wide
population can be estimated as follows:
\begin{align}
\mbox{Wide binary fraction}    &=& \binfrac(1-\binfrac_0)^2           \\
\mbox{Wide triple fraction}    &=& 2 \binfrac\binfrac_0(1-\binfrac_0)  \\
\mbox{Wide quadruple fraction} &=& \binfrac\binfrac_0^2 \,,               
\end{align}
where we have made the assumption that none of the primordial binary
systems has broken up.

All models shown in Fig.~\ref{figure:binaritysma} result in wide
binary fractions $\binfrac \approx 8\%$ that are more or less
independent of $\binfrac_0$ and $a_0$. The value of $\binfrac$ is
therefore primarily determined by the initial values of the number of
system $N$ in the cluster, and its initial size $R$. 

If we assume that the wide orbits in the bottom panels of
Fig.~\ref{figure:binaritysma} (where $\binfrac_0=50\%$) are formed of
randomly paired components (i.e., single stars or primordial
binaries), we can calculate the fraction of higher-order multiple systems among the
$\binfrac=8\%$ wide binaries. Among these, we predict that $25\%$,
$50\%$, and $25\%$, are binary, triple, and quadruple systems,
respectively. In this example, we thus expect 75\% of the ``wide
binaries'' to be higher-order multiple systems.  Due to the random process, the
outer orbits of these systems are expected to be uncorrelated with the
inner orbits or stellar spin axes.

The ratios between wide binary, triple, and quadruple systems are
therefore indicative of $\binfrac_0$. A survey for higher-order
multiplicity among ``wide binary systems'' can thus be used to
constrain the primordial binary fraction. Given the fact that the
majority of stars do form in binary systems, we predict a very
high fraction of higher-order multiple systems among wide ``binary'' systems; see
Fig.~\ref{figure:multiples}. Our proposed mechanism could explain the
existence of the observed wide multiple systems \citep[][]{mamajek2009},
and our predictions are strongly supported by the
surveys of \cite{makarov2008} and \cite{faherty2010}, who find that a
high fraction of the common proper motion pairs in their survey contain inner binary
or triple systems, which is significantly higher than in populations
of other types of binaries.

\begin{figure}
  \centering
  \includegraphics[width=0.5\textwidth,height=!]{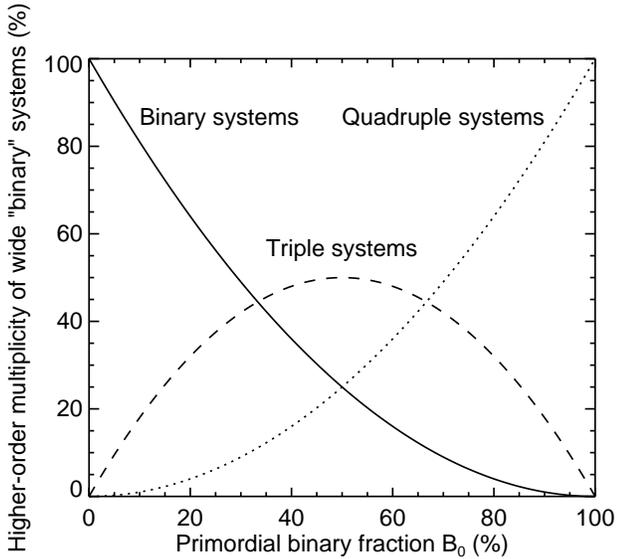}
  \caption{Given the fact that most star form in binary systems, it is
  expected that the majority of the wide ``binary'' systems are in
  fact part of a triple or quadruple system.
  \label{figure:multiples} }
\end{figure}


\section{Summary and discussion}\label{section:conclusions}

Observations have shown that $\sim 15\%$ of binaries are wide ($a>
10^3$~au). These wide binaries are difficult to explain as being the
result of star formation as it is difficult to see how wide binaries
can form (especially those $> 10^4$~au), and they would be rapidly
destroyed in clustered star forming environments.  Whilst $10$ --
$30\%$ of stars do appear to form in an `isolated' environment in
which such binaries could possibly survive, in order to explain the
fraction of wide binaries, almost all stars forming in isolated
environments would have to form wide binaries.  Further, such wide
binaries cannot be formed later in any significant numbers by
dynamical interactions in the Galactic field.

In this paper we study the possibility of wide binary formation during
the dissolution phase of star clusters, in particular, during the
rapid expansion of clusters after gas expulsion.  We study this
possibility using (1) an analytical approach in an idealised
situation, (2) a Monte Carlo approach, and (3) detailed $N$-body
simulations. Our main conclusions are as follows:
\begin{enumerate}

\item The wide binary fraction $\binfrac$ among the dissolved stellar
      population ranges between 1\% and 30\%, depending on the cluster
      properties. 

\item More massive star clusters result in a smaller wide $\binfrac$ than
      low-mass clusters. Clusters with a spherical, smooth stellar
      density distribution form fewer wide binaries than substructured
      clusters of the same size and mass. This is due to the fact that
      the average distance between nearest neighbours is smaller for
      substructured clusters. Expanding (post-gas expulsion) star
      clusters produce a larger $\binfrac$ than those starting
      out of equilibrium.

\item The typical semi-major axis $a$ of the newly formed binaries is
      similar to the initial size $R$ of the star cluster from which 
      they were born. The resulting semi-major axis distribution is
      generally bimodal, consisting of a \dynamicalpeak{} with binary
      systems formed by dynamical interactions, and a
      \dissolutionpeak{} with binary systems formed during the cluster
      dissolution phase.

\item The formation of wide binaries during the star cluster
      dissolution phase is a random process, resulting in the
      following orbital properties. The eccentricity distribution of
      the wide binaries is approximately thermal: $f(e) \approx 2e$
      for $0\leq e < 1$. The mass ratio distribution of the wide
      binaries is the result of gravitationally-focused random
      pairing. In a wide binary, the orbital and spin angular momenta
      are uncorrelated.

    \item Star clusters with a non-zero primordial binary population
      form wide triple and quadruple systems, i.e., the components of
      a newly-formed wide ``binary'' can themselves be close
      primordial binaries, rather than single stars.  The ratio of
      triple to quadruple systems among very wide orbits is therefore
      indicative of the primordial binary fraction
      $\binfrac_0$. Given that $\binfrac_0$ is large, we predict a
      high frequency of triple and quadruple systems among the known
      wide ``binary'' systems, which is supported by existing surveys
      for higher-order multiplicity among wide binary systems.

\end{enumerate}

Throughout this paper we have made predictions of the properties of
the wide binary population resulting from the dissolution of
individual clusters. In order to compare our results with
observations, we should therefore take into account the fact that the
field star population is made up of the stars resulting from an
ensemble of clusters of different sizes and masses. The initial
cluster mass distribution may be approximated by $f(M) \propto
M^{-\gamma}$ with $\gamma \approx 2$ \citep[see, e.g.,][]{zhang1999,
  ashman2001, bik2003, hunter2003}. Given the number of stars
$N=\mcl/\langle m \rangle$, where $\langle m \rangle$ is the average
mass of a star, this distribution is equivalent to $f(N) \propto
N^{-\gamma}$. \cite{oey2004} suggest that the above expression can be
extrapolated down to $N_{\rm min}=1$. The upper limit for the initial
cluster mass distribution is $M_{\rm max} \approx 10^6\msun$
\citep[e.g.,][and references therein]{degrijsparmentier2007}. The
resulting binary fraction $\binfrac_f$ for the ensemble of stars
(i.e., the field star population) is then given by:
\begin{equation}
  \binfrac_f = \frac{ \int_{N_{\rm min}}^{N_{\rm max}} \binfrac(\ncl) N
  f(\ncl) d\ncl }{ \int_{N_{\rm min}}^{N_{\rm max}} f(\ncl) N d\ncl }
  \,,
\end{equation}
where $\binfrac(\ncl)$ is the cluster mass dependent wide binary
fraction. The numerator in the above expression is proportional to the
number of binaries, and the denominator is proportional to the total
number of stars in the ensemble of clusters. In addition, the size and
dissolution time of a star cluster, and therefore the wide binary
fraction, may also depend on its Galactic location
\citep[e.g.,][]{baumgardt2003}. An inspection of
Fig.~\ref{figure:nr_nbody} shows that an extrapolation to $N \approx
10^6$ results in a wide binary fraction of several per cent; 
smaller than the observed 15\%, irrespective of the choices for $R$,
$Q$, and the morphology of the cluster. Although we predict rather
small values, our back-of-the-envelope calculation does result in the
right order of magnitude for the wide binary fraction in the Galactic
field.  It is clear, however, that a deeper investigation is required
to accurately recover the properties of the wide binary population in
the field. In particular, a wider range of star cluster morphologies
has to be considered, by varying the fractal dimension
and position-velocity correlations of individual star clusters.

Our proposed formation mechanism for very wide binaries predicts at
least several common proper motion pairs in and around dissolving star
clusters and moving groups. For example, the mechanism may well
explain the presence of the three common proper motion pairs in the
moving groups studied by \cite{clarke2009}. The future prospects in
wide binary research are bright: an enormous number of wide binaries
are expected to be found with the GAIA
mission\footnote{http://www.esa.int/science/gaia} \citep{perryman2001,
turon2005} and LAMOST
\footnote{http://www.lamost.org} \citep{chu1998, stone2008}.  These
datasets should help determine the true fraction of wide binaries and
their orbital parameters.


\section*{Acknowledgements}
We would like to thank Anthony Whitworth (our referee) and Simon
Portegies Zwart for a useful discussion on this topic.  MBNK was
supported by the Peter and Patricia Gruber Foundation through the PPGF
fellowship, the Peking University One Hundred Talent Fund (985), and
by PPARC/STFC (grant PP/D002036/1). We would like to acknowledge CICS
for the provision of research computing facilities through the White
Rose Grid. We acknowledge research support from and hospitality at the
International Space Science Institute in Bern (Switzerland), as part
of an International Team programme. The authors acknowledge the
Sheffield-Bonn Royal Society International Joint Project grant, which
provided financial support and the collaborative opportunities for
this work. The calculations performed by DM were carried out on
computer hardware purchased from the Royal Physiographic Society in
Lund. RJP acknowledges financial support from STFC.


\bibliographystyle{mn2e}
\bibliography{bibliography}


\bsp

\label{lastpage}

\end{document}